\documentclass[reprint,amsmath,amssymb,pre,twocolumn]{revtex4-2}
\usepackage[utf8]{inputenc}

\usepackage{bm}
\usepackage{graphicx}
\usepackage{upgreek}
\usepackage{amsmath}
\usepackage{amssymb}
\usepackage{numprint}
\usepackage{xspace}
\usepackage{comment}
\usepackage[normalem]{ulem}
\usepackage[dvipsnames]{xcolor}

\definecolor{myred}{rgb}{0.78, 0, 0}
\definecolor{myblue}{rgb}{0, 0, 0.78}

\usepackage{hyperref}
\hypersetup{colorlinks = true, citecolor = myblue, linkcolor = myblue}

\newcommand{\kB}{k_\mathrm{B}}

\newcommand{\Eqref}[1]{Eq.~\ref{#1}}

\newcommand{\figref}[1]{Fig.~\ref{#1}}

\DeclareMathOperator{\arctanh}{arctanh}

\newcommand{\phiS}{\phi_\mathrm{s}}
\newcommand{\barphiS}{\bar{\phi}_\mathrm{s}}

\newcommand{\bT}{b_\mathrm{T}}
\newcommand{\bB}{b_\mathrm{B}}
\newcommand{\sbT}{\sigma_\mathrm{T}}
\newcommand{\sbB}{\sigma_\mathrm{B}}
\newcommand{\MA}{M_\mathrm{All}}
\newcommand{\MB}{M_\mathrm{Binary}}
\newcommand{\MT}{M_\mathrm{Ternary}}
\newcommand{\chiEff}{\chi^\mathrm{eff}}

\newcommand{\sbo}{\sigma_\mathrm{o}}
\newcommand{\sbs}{\sigma_\mathrm{s}}

\newcommand{\chiB}{\chi^\mathrm{B}}
\newcommand{\bo}{b_\mathrm{o}}
\newcommand{\bs}{b_\mathrm{s}}
\graphicspath{{./figure_for_paper}}

\begin{document}

\title{Beyond Pairwise: Higher-order physical interactions affect phase separation in multi-component liquids}

\author{Chengjie Luo%
  \email{chengjie.luo@ds.mpg.de}}
\author{Yicheng Qiang%
  \email{yicheng.qiang@ds.mpg.de}}
\author{David Zwicker%
  \email{david.zwicker@ds.mpg.de}}
\affiliation{Max Planck Institute for Dynamics and Self-Organization, Am Faßberg 17,
  37077 Göttingen, Germany}

\begin{abstract}
Phase separation, crucial for spatially segregating biomolecules in cells, is well-understood in the simple case of a few components with pairwise interactions. Yet, biological cells challenge the simple picture in at least two ways: First, biomolecules, like proteins and nucleic acids, exhibit complex, higher-order interactions, where a single molecule may interact with multiple others simultaneously. Second, cells comprise a myriad of different components that form various droplets. Such multicomponent phase separation has been studied in the simple case of pairwise interactions, but an analysis of higher-order interactions is lacking. We propose such a theory and study the corresponding phase diagrams numerically. We find that interactions between three components are similar to pairwise interactions, whereas composition-dependent higher-order interactions between two components can oppose phase separation. This surprising result can only be revealed from the equilibrium phase diagrams, implying that the often-used stability analysis of homogeneous states is inadequate to study these systems. We thus show that higher-order interactions could play a crucial role in forming droplets in cells, and their manipulation could offer novel approaches to controlling multicomponent phase separation.
\end{abstract}

\maketitle
\section{Introduction}
Phase separation provides a thermodynamic mechanism that partitions multicomponent liquids.
The spontaneous segregation of molecules into droplets is crucial in chemical engineering and is implicated in the intricate organization of biological cells.
Biological cells comprise thousands of different components that separate into various droplets known as biomolecular condensates~\cite{banani2017biomolecular}, each containing hundreds of components~\cite{Youn2019}.
Yet, the formation of condensates is often discussed in the simple framework of a binary Flory-Huggins theory~\cite{Hyman2014,julicher2023droplet}, which describes how two components segregate from each other because of an effective repulsive interaction~\cite{Flory1942, huggins1941solutions}.
While this simple theory qualitatively explains phase separation of a single condensate type, it does not account for the complexity in cells.
A defining feature of cells is that they are alive, so phase separation might be affected by active processes~\cite{zwicker2022intertwined}.
However, even the passive behavior of biomolecules in cells differs from the simplest model:
First, biomolecules have complex interactions that are not captured by the Flory-Huggins free energy, and, second, there are thousands of different molecules that form multiple types of condensates simultaneously.
While these challenges have been addressed separately, a combined theory describing both aspects is lacking.

The molecular interactions of biomolecules, like proteins and nucleic acids, are difficult to measure experimentally since they are generally weak and depend on the physiochemical properties of their surrounding~\cite{dignon2020biomolecular}.
Partial experimental measurements can be used to inform theory~\cite{doi1996introduction,fredrickson2006equilibrium, riback2020composition,arter2022biomolecular,saar2023theoretical}, and coarse-grained simulations can unveil sequence-based grammars to describe phase separation~\cite{holehouse2023molecular,lin2018theories,pappu2023phase}.
These works, and even simple polymer models~\cite{dudowicz2002FloryHuggins,pesci1989Lattice}, suggest that interaction energies of biomolecules cannot be characterized by the simple quadratic terms in the Flory-Huggins theory~\cite{Flory1942, huggins1941solutions}.
This is not surprising given the complexity of the molecules and the high volume fraction of macromolecules of about $40\,\%$ in cells~\cite{sear2005cytoplasm}, which together suggest that a virial expansion only up to second order is insufficient~\cite{bruns1997third}.
Moreover, biomolecules tend to have many different domains~\cite{choi2020physical, mohantycomplex}, so they can interact with multiple components simultaneously.
Such higher-order interactions appear naturally in colloidal systems~\cite{russ2002three,brunner2004direct,dobnikar2004three}, lead to complex competitions between components~\cite{sanders2020competing}, enable allosteric effects~\cite{mccullagh2024allosteric}, and explain the behavior of ternary lipid systems~\cite{idema2009phase}.
More broadly, higher-order interactions are vital in ecological networks~\cite{billick1994higher,mayfield2017higher,battiston2020networks,grilli2017higher,bairey2016high,gibbs2022coexistence,kleinhesselink2022detecting,li2020advances}, and they generally emerge from coarse-graining complex interactions~\cite{thibeault2024low, PRXLife.1.023012}.
It is thus likely that higher-order interactions are also crucial to explain the observed complex phase diagrams in cells~\cite{riback2020composition}.

Another complexity of cellular phase separation originates from the many different components that form multiple distinct droplets.
Such multicomponent phase separation currently cannot be described adequately with detailed numerical simulations, so more abstract approaches have been developed~\cite{jacobs2023theory}.
However, even for the relatively simple Flory-Huggins theory, concrete phase diagrams are complex and can only be predicted for few components~\cite{Mao2018, jacobs2023theory}.
Liquids with a large number of components have been studied using linear stability of the homogeneous state.
For random pairwise interactions, this analysis predicts that multicomponent liquids either demix into as many phases as components, or condense into just one or two phases~\cite{Sear2003, sear2005cytoplasm, Jacobs2017, Shrinivas2021}.
Since neither behavior describes biological reality, this suggests that either linear stability analysis is inadequate or higher-order interactions are relevant in cellular phase separation.

To understand the impact of higher-order interactions on multicomponent phase separation, we study a systematically extended Flory-Huggins theory.
We find that interactions that involve three different species play a similar role to pairwise interactions, whereas higher-order interactions that only involve two species, describing composition-dependent interactions, can strongly oppose normal tendencies to phase separate.
Some of these qualitative differences can only be revealed from equilibrium phase diagrams, but not from the simpler linear stability analysis.

\section{Theory}
\subsection{Multicomponent mixtures with cubic interactions}
We consider an incompressible, isothermal mixture of $N$ species, including one inert solvent~$S$.
The composition is then described by the volume fractions $\phi_i$ of the $N-1$ interacting species, while the solvent fraction is $\phiS=1-\sum_{i=1}^{N-1}{\phi_i}$.
The thermodynamics of the system are governed by the free energy density
\begin{multline}
  \label{eq:free_energy}
  f(\phi_1,\hdots,\phi_{N-1}) = \frac{\kB T}{\nu} \Biggl[
   \phiS\ln\phiS + \sum_{i=1}^{N-1}\phi_i\ln \phi_i
\\ 
  	+ \sum_{i,j=1}^{N-1}\frac{\chi_{ij}}{2}\phi_i\phi_j
	+\sum_{i,j,k=1}^{N-1}\frac{b_{ijk}}{3}\phi_i\phi_j\phi_k
   \Biggr]
   \;,
\end{multline} 
where $\kB T$ is the relevant energy scale, and $\nu$ denotes a molecular volume, which we consider to be the same for all species for simplicity.
The terms on the first line capture the translational entropy of all species, while all other contributions to the energy are described by the terms in the second line. 
The second line can be interpreted as a Taylor expansion in terms of the densities $\{\phi_i\}$. 
The lowest orders of the expansion do not contribute since the zeroth-order term just shifts the overall energy and terms linear in $\phi_i$ drop out in the equilibrium conditions; see below.
The first relevant term is the quadratic one, which quantifies the principle pair interactions between species $i$ and $j$ by Flory parameters $\chi_{ij}$; positive values denote effective repulsion of species $i$ and $j$, whereas negative values lead to attraction~\cite{Flory1942, huggins1941solutions}.
Finally, the last term in \Eqref{eq:free_energy} quantifies cubic interactions among species $i$, $j$, and $k$ by a parameter $b_{ijk}$, which we address in this paper.
Terms of higher order could be included (see Appendix for a general framework), but we limit the discussion to third-order interactions for simplicity.
The sums in \Eqref{eq:free_energy} imply that only the symmetric parts of the interaction parameters contribute, so we can assume $\chi_{ij}=\chi_{ji}$ and $b_{ijk}=b_{ikj}=b_{jik}=b_{jki}=b_{kij}=b_{kji}$. 
Moreover, we exploit incompressibility ($\phi_i=1-\sum_{j\neq i}\phi_j$), and the fact that terms linear in $\phi_i$ do not change the equilibrium states, to  remove diagonal entries ($\chi_{ii}=b_{iii}=0$).
The thermodynamics of the liquid are then fully described by the two sets of interaction parameters $\chi_{ij}$ and $b_{ijk}$.

The cubic interactions quantified by~$b_{ijk}$ describe two fundamentally different physical processes.
To see this, we rewrite the second line in \Eqref{eq:free_energy} as $\frac12 \sum_{i,j}\chiEff_{ij}\phi_i\phi_j$ by introducing re-scaled pair interactions
\begin{equation}
  \label{eq:effective_chi}
  \chiEff_{ij}= %(\{\phi_i\}) =
  	\chi_{ij} 
	+  \underbrace{ \tfrac23(b_{iji}\phi_i+b_{ijj}\phi_j)}_\text{\hspace{-1cm}binary cubic interaction\hspace{-1cm}}
	+ \underbrace{\tfrac23\hspace{-2mm}\sum_{k\neq i, k\neq j}\hspace{-2mm}b_{ijk}\phi_k}_\text{\hspace{-1cm}ternary cubic interaction\hspace{-1cm}}
	\;,
\end{equation}
which now depend on the composition $\{\phi_i\}$.
The first term on the right hand side summarizes the familiar quadratic interactions, but we split the cubic interactions into two fundamentally different contributions:
The last term captures true cubic interactions between three different species ($i\neq j \neq k$), which we thus call \emph{ternary cubic interactions}.
In contrast, the middle term describes composition-dependent pair interactions, where the coefficients $b_{iji}$ and $b_{ijj}$ capture how the pair-interaction $\chi^\mathrm{eff}_{ij}$ depends on the composition of the two involved species, and we thus call these \emph{binary cubic interactions}.
Such composition-dependent pair interactions are often used for describing real polymers~\cite{baulin2002concentration}.

In a system with $N$ components (including the inert solvent), the interaction parameters $\chi_{ij}$ and $b_{ijk}$ have $\frac12(N-1)(N-2)$ and $\frac16(N-1)(N-2)(N-3)$ independent entries, respectively.
To limit this large parameter space, we for simplicity only consider random interaction matrices, where the entries of $\chi_{ij}$ and $b_{ijk}$ are drawn independently from normal distributions.
Specifically, we chose the quadratic interactions as $\chi_{ij}\sim \mathcal{N}(\chi,\sigma_{\chi}^2)$, the binary cubic interactions as $b_{iij}\sim \mathcal{N}(\bB,\sbB^2)$, and  the ternary cubic interactions as $b_{ijk}\sim \mathcal{N}(\bT,\sbT^2)$, where $\mathcal{N}(\mu,\sigma^2)$ represents a normal distribution with mean $\mu$ and variance $\sigma^2$.
Our model is thus parametrized by the component count~$N$ and six parameters describing the distributions of the interaction parameters.

\subsection{Unstable modes of homogeneous states}
One approach to investigate the behavior of multicomponent liquids is to analyze homogeneous states, which are quantified by the volume fractions $\{\phi_1,\hdots,\phi_{N-1}\}$.
Such a state is stable if the Hessian of the free energy,
\begin{equation}
  %H_{ij}=
  \frac{\partial^2{f}}{\partial{\phi_i}\partial{\phi_j}}
  =  \frac{\delta_{ij}}{\phi_i} + \frac{1}{\phiS} + \chi_{ij} + 2\sum_{k=1}^{N-1}b_{ijk}\phi_k
  \;,
\end{equation}
is positive definite, i.e., if all its eigenvalues are positive.
In contrast, the homogeneous state is unstable if one of the eigenvalues is negative, indicating that the system would split into different phases.
We expect that the number of unstable modes~$U$, i.e., the number of negative eigenvalues, is correlated with the number~$M$ of distinct phases that form during phase separation.
For instance,  the homogeneous state must be stable when it is the only stable state ($M=1$ implies $U=0$).
Moreover, we expect the maximal number of phases ($M=N+1$) when all modes are unstable ($U=N$).
Taken together, the simplest hypothesis for the correlation is thus $M=U+1$, but since the eigenvalue analysis strictly only applies to infinitesimal small deviations from the homogeneous phase, it is unclear how well $U$ predicts fully phase separated states.
To investigate this in detail, we thus also quantify coexisting phases directly.

\subsection{Phase count in equilibrium states}
Coexisting phases need to obey thermodynamic conditions, enforcing equal chemical potentials and pressures in all phases~\cite{zwicker2022intertwined}.
Denoting the volume fraction of the $i$-th species in the $n$-th phase by $\phi_i^{(n)}$, we express these intensive quantities in non-dimensional form, $\mu_i^{(n)}=\nu / (\kB T) \partial f/\partial \phi_i^{(n)}$ and $P^{(n)}=\nu/\kB T[\phi_i^{(n)}\partial f/\partial \phi_i^{(n)}-f]$, implying
\begin{subequations}
\begin{align}
 \mu_i^{(n)}&=\!
  \sum_{j=1}^{N-1}\chi_{ij}\phi_j^{(n)}+\sum_{j,k=1}^{N-1}b_{ijk}\phi_j^{(n)}\phi_k^{(n)}+\ln\frac{\phi_i^{(n)}}{\phiS^{(n)}}
\\[4pt]
  P^{(n)}&=\!\!
  	\sum_{i,j=1}^{N-1}\frac{\chi_{ij}}{2}\phi_i^{(n)}\phi_j^{(n)}
	+\!\!\!\!\sum_{i,j,k=1}^{N-1}\!\!\frac{2b_{ijk}}{3}\phi_i^{(n)}\phi_j^{(n)}\phi_k^{(n)}-\ln\phiS^{(n)}
%  \;.
\end{align}
\end{subequations}
The coexistence conditions thus read
\begin{align}
\label{eqn:coexistence_conditions}
  \mu_i^{(n)}&=\mu_i^{(m)}
  &&\text{and}&
  P^{(n)} &= P^{(m)}
  \;,
\end{align} 
for $i=1,\ldots, N-1$ and $n,m=1,\ldots,M$. 
These are $N(M-1)$ independent conditions, which need to be satisfied for $M$ phases to coexist.
These non-linear equations are difficult to solve explicitly and we thus resort to a numerical Monte-Carlo method~\cite{zwicker2022evolved} to discover solutions.
Since the original implementation weighted compositions with similar fractions much more heavily~\cite{Jacobs2023}, we have improved the method to find a solution of these equations for a given average fraction.
Briefly, we initialize many compartments with random compositions such that the composition averaged over all compartments assumes a desired value.
We then minimize the overall free energy by exchanging components and volume between compartments until a stationary state is reached, which corresponds to thermodynamic equilibrium~\cite{zwicker2022evolved}. 
We can then analyze the coexisting equilibrium phases  and uniformly sample the phase space.

\section{Results}

To analyze the impact of cubic interactions on phase separation, we will first investigate the simpler situation of few components with symmetric interactions.
After increasing the component count in the third subsection, we will finally study the impact of diverse cubic interactions.

\begin{figure*}
  \centering
  \includegraphics[width=1\textwidth]{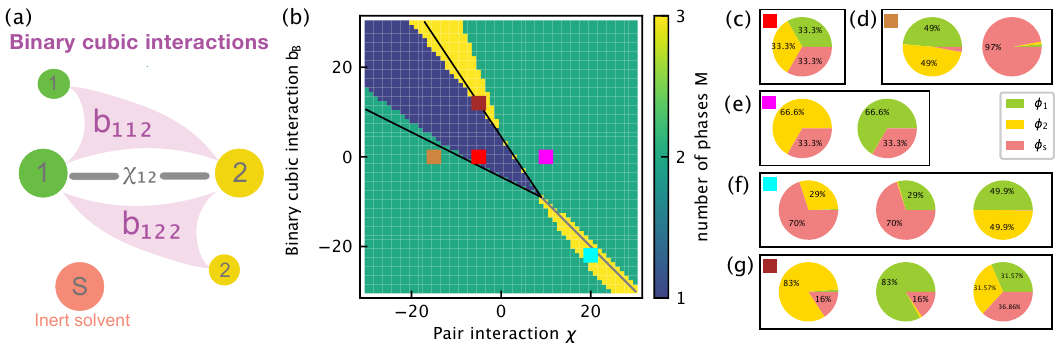}
  \caption{ 
  \textbf{Binary cubic interactions lead to additional phases for symmetric interactions.}
  (a)~Schematic of physical pair interactions $\chi_{12}$ and binary cubic interactions $b_{112}$ and $b_{221}$ of two species $1$, $2$, and the inert solvent $S$.
  (b) Phase count~$M$ as a function of the pairwise interaction strength~$\chi$ and the strength~$\bB$ of the binary cubic  interaction for the symmetric composition $\bar{\phi}_1=\bar{\phi}_2=\barphiS=\frac13$.
  Linear stability analysis predicts a stable homogeneous state in the region enclosed by the solid black line; see Appendix. The grey line represents $\chi=-\bB$, where species $1$, $2$, and the solvent are all equivalent.
  (c)--(g) Compositions of coexisting phases at five parameter values indicated by the colors in panel (b).
  Each pie chart indicates the fractions of the three components in a single phase.
  (b)--(g) Additional model parameters are $N=3$, $\bT=0$ and $\sigma_{\chi}=\sbT=\sbB=0$.
  }
  \label{fig:compositions_three_species} 
\end{figure*}

\subsection{Attractive binary cubic interactions promote phase separation}
We first focus on the effect of the binary cubic interactions in the simplest system with $N=3$ components; see \figref{fig:compositions_three_species}(a).
Since the solvent is inert, only two species interact with each other and the ternary cubic interaction does not contribute.
For simplicity, we first consider identical interactions ($\chi_{12} = \chi$ and $b_{112} = b_{122} = \bB$) and equal fractions of all components ($\bar{\phi}_1=\bar{\phi}_2=\bar\phiS=\frac13$).
\figref{fig:compositions_three_species}(b) reveals that the system exhibits either one or two phases without cubic interactions ($\bB=0$).
The former case corresponds to a stable homogeneous state, see \figref{fig:compositions_three_species}(c), whereas the latter comes in two flavors:
For strong attraction ($\chi<-24\arctanh(1/3)\approx-8.3$),  species $1$ and $2$ are equally enriched in one phase and together segregate from the solvent; see \figref{fig:compositions_three_species}(d).
In contrast, for strong repulsion ($\chi>3$), species $1$ and $2$ separate from each other with equal solvent fractions throughout; see \figref{fig:compositions_three_species}(e).
Binary cubic interactions ($\bB\neq0$) significantly impact the phase diagram, revealing unexpected regions supporting three phases (yellow areas in \figref{fig:compositions_three_species}(b)).
We again observe two qualitatively different regions: 
For negative $\bB$ and positive $\chi$, the two solutes form one phase with barely any solvent, whereas the other two phases exhibit a lot of solvent and either one of the solutes; see \figref{fig:compositions_three_species}(f).
Conversely, for positive $\bB$ and negative $\chi$, each solute dominates in one phase, whereas the third phase has a balanced composition of all components; see \figref{fig:compositions_three_species}(g).
Notably, when $\chi=-\bB$, marked by the gray line in \figref{fig:compositions_three_species}(b), all three species behave equivalently, so the system necessarily forms three phases when the interactions are sufficiently repulsive.
The apparent interaction of the solutes with the inert solvent is a consequence of incompressibility; see Appendix.
This particular case also reveals that the effect of binary cubic interactions can at least partly be described by effective interactions with the inert solvent.
This initial investigation shows that binary cubic interactions $\bB$ affect what phases coexist when they oppose the binary quadratic interactions $\chi$.

\begin{figure*}
  \centering
  \includegraphics[width=1.0\textwidth]{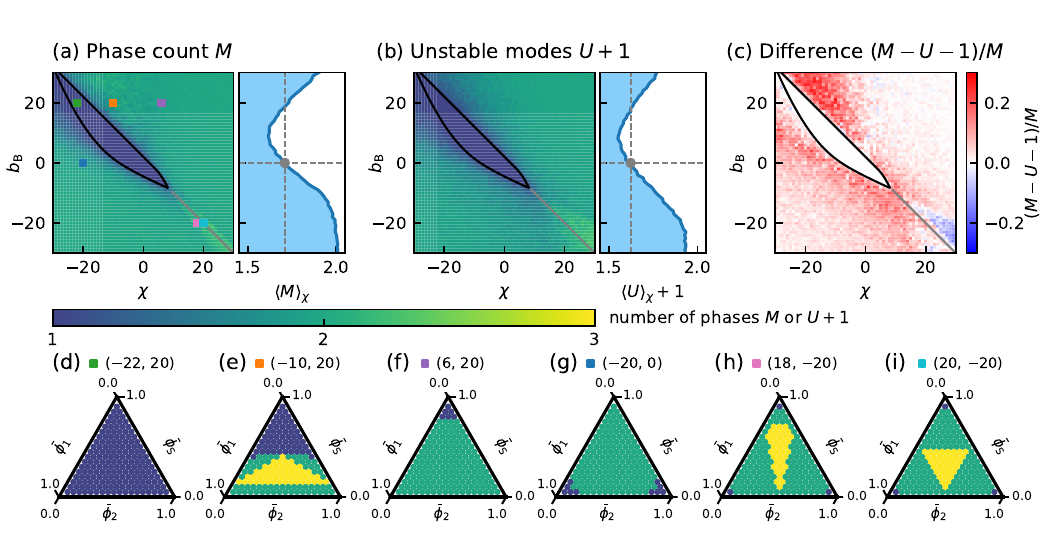}
  \caption{ 
  \textbf{Attractive binary cubic interactions promote phase separation.}
  Results for $N=3$ with equal binary cubic interactions ($\sigma_{\chi}=\sbB=0$).
  (a) Mean phase count $M$ as a function of the pairwise interaction strength~$\chi$ and the strength~$\bB$ of the binary cubic interactions averaged over all compositions.
  The right panel shows $M$ additionally averaged over $\chi$; the dashed line marks the value at $\bB=0$.
  (b) Mean number of unstable modes $U$ as a function of $\chi$ and $\bB$.
  The right panel shows $U$ averaged over $\chi$; the dashed line marks the value at $\bB=0$.
  (c) Relative difference $(M-U-1)/M$ as a function of $\chi$ and $\bB$.
  (a)--(c) The black curve is the boundary of the homogeneous stable region in which no phase separation can happen for any mean volume fractions.
  $100$ compositions have been sampled uniformly for each set of parameter values.
  (d)--(i) Phase diagrams as a function of the fractions $\bar{\phi}_1$ and $\bar{\phi}_2$ of the interacting species $1$ and $2$ for the indicated parameters $(\chi,\bB)$.
}
  \label{fig:phase_diagrams_three_species} 
\end{figure*}

To analyze the effect of binary cubic interactions on the three-component system further, we next investigate the phase diagrams as a function of compositions at fixed $\chi$ and $\bB$.
\figref{fig:phase_diagrams_three_species}(a) shows the phase count~$M$ averaged over all possible compositions.
Most of the phase diagram exhibits roughly two phases, but the homogeneous state dominates in a part of the upper left quadrant; similar to \figref{fig:compositions_three_species}(b).
Interestingly, the two parameter regions that exhibited three phases for equal composition (see \figref{fig:compositions_three_species}(b)) are now less prominent.
To analyze the effect of the cubic interactions more clearly, we average $M$ over all values of $\chi$; see right panel in \figref{fig:phase_diagrams_three_species}(a).
This quantification reveals that attractive binary cubic interactions ($\bB < 0$) increase the average phase count and thus promote phase separation, whereas repulsive interactions tend to suppress phase separation unless they are very strong.
A similar result is also visible in the quantification of the number~$U$ of unstable modes shown in \figref{fig:phase_diagrams_three_species}(b), although the direct comparison between $U$ and $M$ shown in  \figref{fig:phase_diagrams_three_species}(c) reveals that $U$ often deviates substantially from $M$.

To understand the role of composition in more detail, we next determine the full phase diagrams for particular choices of $(\chi, \bB)$; see \mbox{\figref{fig:phase_diagrams_three_species}(d--i)}.
For large positive $\bB$, the homogenous phase is the only stable phase when $\chi \approx -\bB$ (\figref{fig:phase_diagrams_three_species}(d)), indicating that repulsive binary cubic interactions can suppress phase separation.
However, slightly larger values of $\chi$ can lead to rich phase diagrams, also involving three-phase regions (\figref{fig:phase_diagrams_three_species}(e)), until the two-phase region dominates for positive $\chi$ (\figref{fig:phase_diagrams_three_species}(f)). 
Here, the homogeneous system is only favored when the solvent fraction~$\barphiS$ is very high.
Conversely, the homogeneous state is favored for low $\barphiS$ when the cubic interactions are absent; see \figref{fig:phase_diagrams_three_species}(g).
When the binary cubic interactions are attractive ($\bB<0$), the homogenous state is rare and we again find three-phase regions; see \figref{fig:phase_diagrams_three_species}(h).
Interestingly, the width of the three-phase region becomes smaller for small~$\barphiS$, whereas it became wider in \figref{fig:phase_diagrams_three_species}(e), which might be caused by the opposite signs of the interactions in these two cases.
Moreover, \figref{fig:phase_diagrams_three_species}(i) reveals that the special case of $\chi = -\bB$ indeed leads to symmetric interactions between all species.
stinct roles played by binary cubic attraction and repulsion.

\begin{figure*}
  \centering
  \includegraphics[width=1\textwidth]{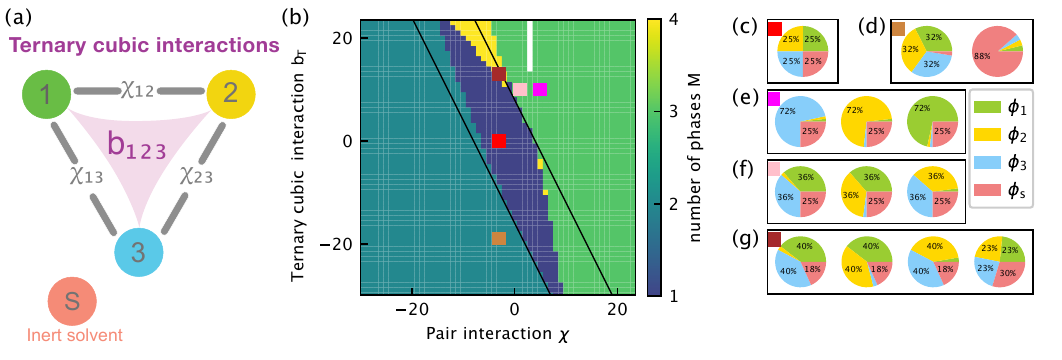}
  \caption{ 
  \textbf{Ternary cubic interactions lead to additional phases for symmetric interactions.}
  (a)~Schematic of physical pair interactions $\chi_{ij}$ and ternary cubic interactions $b_{ijk}$ of three species $1$, $2$, $3$, and the inert solvent $S$.
  (b) Phase count~$M$ as a function of the pairwise interaction strength~$\chi$ and the strength~$\bT$ of the ternary cubic interaction for the symmetric composition $\bar{\phi}_1=\bar{\phi}_2=\bar{\phi}_3=\barphiS=\frac14$.
  Linear stability analysis predicts a stable homogeneous state between the solid black lines; see Appendix.
  The white area exhibits six phases which originate from degeneracies caused by the identical interactions between solutes. 
  (c)--(g) Compositions of coexisting phases at five parameter values indicated by the colors in panel (b).
  Each pie chart indicates the fractions of the four components in a single phase.
  (b)--(g) Additional model parameters are $N=4$, $\bB=0$ and $\sigma_{\chi}=\sbT=\sbB=0$.
  }
  \label{fig:pie} 
\end{figure*}

\begin{figure*}
  \centering
  \includegraphics[width=1.0\textwidth]{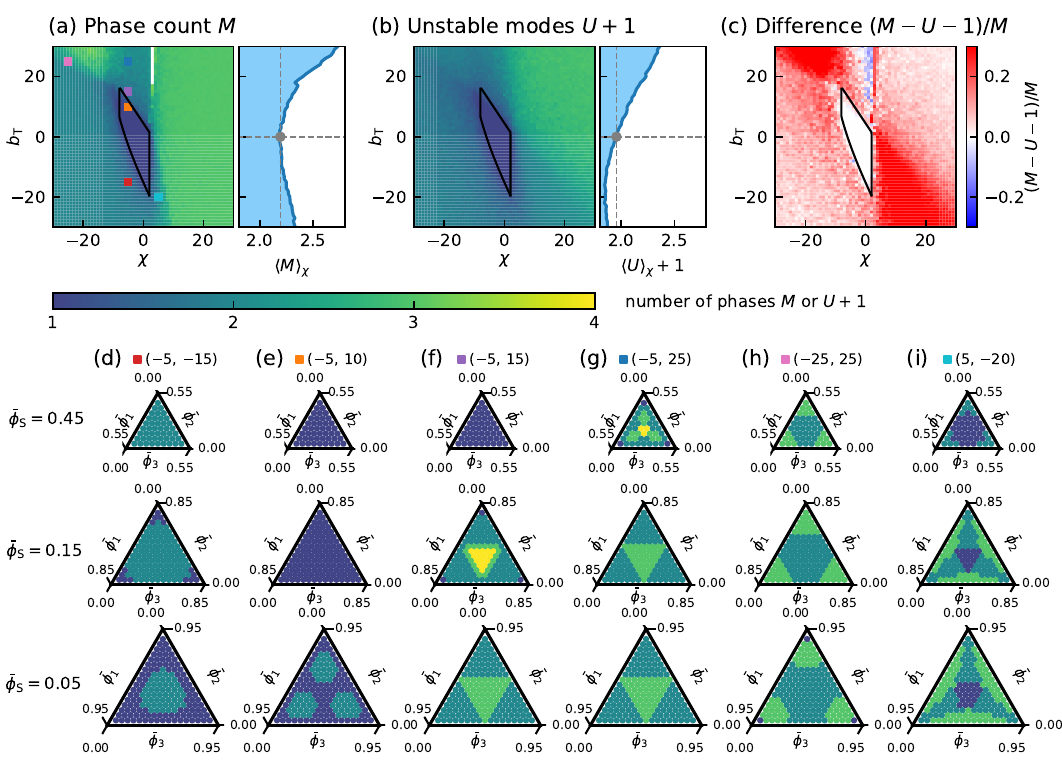}
  \caption{ 
  \textbf{Repulsive ternary cubic interactions promote phase separation.}
  Results for $N=4$ with equal ternary cubic interactions ($\bB=0$ and $\sigma_{\chi}=\sbB= \sbT=0$).
  (a) Mean phase count $M$ as a function of the pairwise interaction strength~$\chi$ and the strength~$\bT$ of the ternary cubic interaction averaged over all compositions.
  The right panel shows $M$ additionally averaged over $\chi$; the dashed line marks the value at $\bT = 0$.
  (b) Mean number of unstable modes $U$ as a function of $\chi$ and $\bT$. 
  The right panel shows $U$ averaged over $\chi$; the dashed line marks the value at $\bT = 0$.
  (c) Relative difference $(M-U-1)/M$ as a function of $\chi$ and $\bT$.
  (a)--(c) The black curve is the boundary of the homogeneous stable region in which no phase separation can happen for any mean volume fractions.
  $100$ compositions have been sampled uniformly for each set of parameter values.
  (d)--(i) Ternary phase diagrams as a function of the fractions $\phi_i$ of the interacting species $i=1,2,3$ for the indicated parameters $(\chi,\bT)$ and various solvent fractions $\barphiS$.
}
  \label{fig:phase_diagrams} 
\end{figure*}

The influence of binary cubic interactions can be partly rationalized by interpreting them as effective pair interactions.
For $N=3$ and $\sbB=0$, \Eqref{eq:effective_chi} reduces to $\chiEff_{12} = \chi + \bB(1 - \phiS)$, implying that repulsive cubic interactions ($\bB>0$) generally increase $\chiEff_{12}$.
This equation also reveals that the sum of the parameters $\chi$ and $\bB$ is crucial, explaining why interesting phase emerge in the region where $|\chi+\bB|$ is small.
The effect on $\chiEff_{12}$ is particularly large when the fraction $\phiS$ of the inert solvent is low, explaining why we find more phases in the lower part of \figref{fig:phase_diagrams_three_species}(e).
Conversely, for attractive cubic interactions ($\bB<0$), we find $\chiEff_{12} \approx |\bB|\phiS$ in the special case $\chi + \bB \approx 0$, showing that the effective interaction becomes more repulsive for larger solvent fraction~$\phiS$, consistent with \figref{fig:phase_diagrams_three_species}(h).
Taken together, this analysis again reveals that binary cubic interactions can partly be interpreted as interactions with the inert solvent.
In particular, repulsive binary cubic interactions stabilize homogeneous states, whereas attractive interactions promote phase separation.

\subsection{Repulsive ternary cubic interactions promote phase separation}

We next analyze the effect of ternary cubic interactions that characterize the interplay among three different species in a liquid with $N=4$ components; see \figref{fig:pie}(a).
We start by considering identical interactions across all species (vanishing variances; $\sigma_{\chi}^2=\sbB^2=\sbT^2=0$) without binary cubic interactions ($\bB=0$) for equal fractions ($\bar{\phi}_1=\bar{\phi}_2=\bar{\phi}_3=\barphiS=\frac14$), so the only two control parameters are $\chi$ and $b_{123} =\bT$.
\figref{fig:pie}(b) shows the system exhibits between one and three phases without ternary cubic interactions ($\bT=0$).
In particular, only the homogeneous state is stable for weak binary interactions (small $|\chi|$); see \figref{fig:pie}(c).
Strong attraction ($\chi< -6 \ln(3)\approx -6.59$) results in a phase enriched in species $1$, $2$, and $3$, which together segregate from the inert solvent, and hence form two phases whose compositions are similar to that shown in \figref{fig:pie}(d).
In contrast, for strong repulsion ($\chi>4$), species $1$, $2$, and $3$ separate from each other with equal solvent fractions throughout, leading to three phases with compositions similar to those shown in \figref{fig:pie}(e).
Adding ternary cubic interactions ($\bT\neq0$) enriches the phase diagram:
In particular, we now find one parameter region supporting four phases when ternary cubic interactions are repulsive ($\bT>0$, but $\chi<0$); see yellow area in \figref{fig:pie}(b).
Surprisingly, two solute components co-segregate together in three of the phases, whereas the solvent dominates the fourth one; see \figref{fig:pie}(g). 
The detailed compositions also reveal additional transitions in regions where panel b reports the same phase count.
For instance, in the large bright green region ($M=3$), we observe a transition from three phases where always two solutes co-segregate (\figref{fig:pie}(f), $\chi<3$) to three phases that are each dominated by a single solute (\figref{fig:pie}(e), $\chi>3$).
Taken together, this initial analysis suggests ternary cubic interactions result in interesting phases, which cannot be explained by a simple re-scaling of quadratic interactions.

To analyze the effect of composition, we next investigate the phase count $M$ averaged over all compositions. 
The features of the respective diagram shown in \figref{fig:phase_diagrams}(a) are similar to \figref{fig:pie}(b);
Attractive quadratic interactions ($\chi<0$) generally lead to two phases, whereas strong repulsion ($\chi>0$) leads to three phases.
Weak interactions stabilize the homogeneous state, although the parameter region where this is the only possible state (enclosed by the black line) is smaller than in \figref{fig:pie}(b).
Adding ternary cubic interactions has qualitatively similar effects to the case of equal composition shown in \figref{fig:pie}(b):
Repulsive interactions ($\bT>0$)  generally increase the phase count~$M$, which is particularly obvious when $M$ is averaged over all $\chi$; see right panel of \figref{fig:phase_diagrams}(a).
Although the number~$U$ of unstable modes shown in \figref{fig:phase_diagrams}(b) paints a similar picture, the predicted effects are much weaker.
Moreover, \figref{fig:phase_diagrams}(c) reveals that $U+1$ consistently underestimates $M$, particularly when ternary cubic interactions oppose quadratic interactions.
Taken together, this suggests that repulsive ternary cubic interactions promote phase separation, similar to repulsive quadratic interactions, and in stark contrast to binary cubic interactions.

\begin{figure*}
  \centering
  \includegraphics[width=1.0\textwidth]{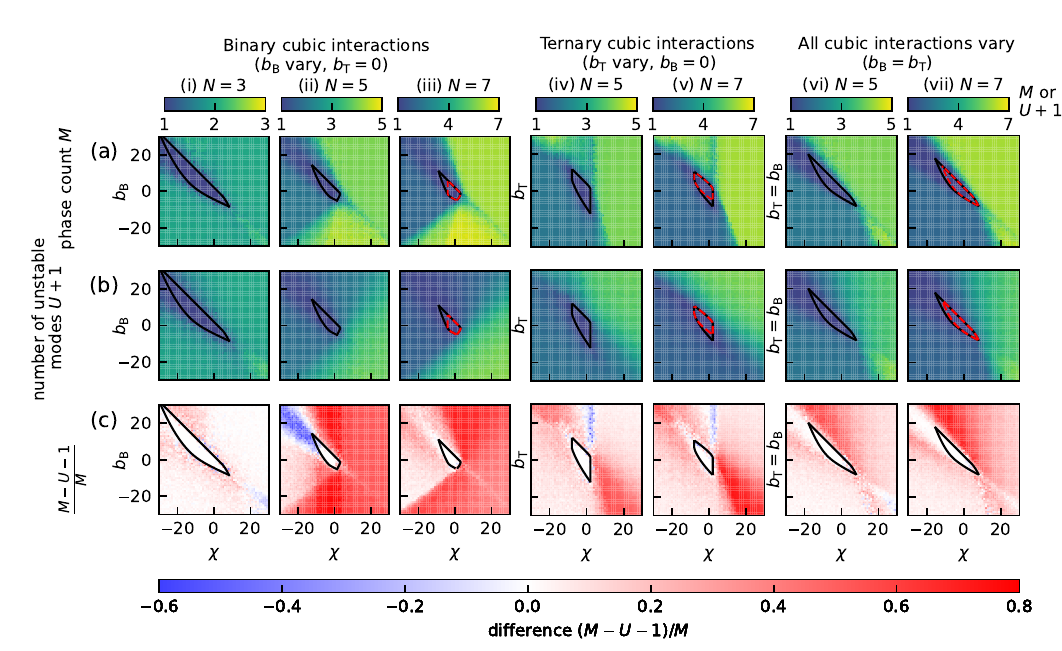}
  \caption{ 
  \textbf{Three-body interactions affect phase count of multicomponent liquids.}
  (a)~Mean phase count $M$ as a function of the pairwise interaction strength~$\chi$ and the strength~$\bB$ of the binary cubic interaction or the strength~$\bT$ of the ternary cubic interaction, averaged over all compositions for $N=3,5,7$.
  (b)~Mean number of unstable modes $U$ as a function of $\chi$ and $\bT$ (or $\bB$) for $N=5,7$.
  (c)~Relative difference $(M-U-1)/M$ as a function of $\chi$ and $\bT$ (or $\bB$) for $N=5,7$.
  (a)--(c) Data involves averages of $100$ uniformly distributed compositions at each point. 
    Linear stability analysis indicates that homogeneous states are stable inside the black line, and the red dashed line marks the limit for $N\rightarrow\infty$; see Appendix.
  (i)--(iii) Binary cubic interactions vary with given $\bB$ and $\sigma_\chi=\sbB=1$, while ternary cubic terms are zero ($\bT=\sbT=0$).
  (iv)--(v) Ternary cubic interactions vary with given $\bT$ and $\sigma_\chi=\sbT=1$, while binary cubic interactions are zero  ($\bB=\sbB=0$).
  (vi)--(vii) All off-diagonal cubic interactions vary with $\bB=\bT$ and $\sigma_\chi=\sbB=\sbT=1$.
  }
  \label{fig:binodal_spinodal_diff_fixed_interaction_random_phi_sigma1} 
\end{figure*}

To understand the role of composition in detail, we next determine the full phase diagram for particular choices of $\chi$, $\bT$, and $\barphiS$; see \mbox{\figref{fig:phase_diagrams}(d--i)}.
%However, interesting phase diagrams appear in other parameter regions; see \figref{fig:phase_diagrams}(d)-(i).
Panel (d) reveals that attractive interactions ($\bT<0$ and $\chi<0$) generally favor two phases, particularly for larger solvent fractions $\barphiS$ (upper row).
In contrast, larger $\barphiS$ hinders the formation of two phases for repulsive ternary cubic interactions (large $\bT$, \figref{fig:phase_diagrams}(e)).
Even stronger ternary  repulsion can facilitate states with four phases (yellow regions in \figref{fig:phase_diagrams}(f)), but both  low and  high $\barphiS$ suppresses this state.
For large positive $\bT$ at low $\barphiS$,  three-phase regions appear for equal composition (bright green regions in lower panel of \figref{fig:phase_diagrams}(g)), while they appear in the corner of the phase diagram when $\chi$ is lower; see \figref{fig:phase_diagrams}(h).
Finally, \figref{fig:phase_diagrams}(i) shows that for positive $\chi$, increasing $\barphiS$ leads to fewer phases, similar to \figref{fig:phase_diagrams}(e), but now three phases can be prevalent.
Taken together, these detailed phase diagrams demonstrate a profound effect of ternary interactions on the actual phases that form for a particular composition of the system.

We demonstrated that ternary cubic interactions affect the number of phases and particularly their composition.
Linear stability analysis predicts this behavior even worse than in the case of binary interactions; compare \figref{fig:phase_diagrams}(c) to \figref{fig:phase_diagrams_three_species}(c).
Although \Eqref{eq:effective_chi} demonstrates that $b_{ijk}$ can be interpreted as composition-dependent pairwise interactions, our numerical results indicate that third-order interactions do not merely rescale the two-body interaction $\chi_{ij}$.
Moreover, binary cubic interactions tend to lead to more phases when they are attractive ($\bB<0$), whereas repulsive ternary cubic interactions ($\bT>0$) tend to increase it.
Taken together, even the simplest cases of $N=3$ and $N=4$ components, respectively, highlight that cubic interactions play a nontrivial role.

\subsection{Ternary cubic interactions dominate binary cubic interactions in case of many components}

\begin{figure}
  \centering
  \includegraphics[width=0.5\textwidth]{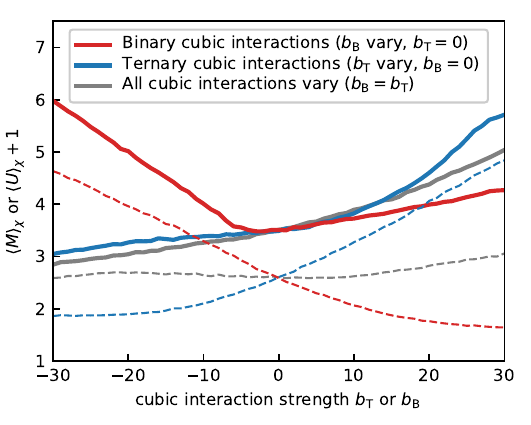}
  \caption{ 
  \textbf{Binary and ternary cubic interactions have opposite effect on phase count. }
  The mean phase count $\langle M \rangle_{\chi}$ (solid curves) and the number of unstable mode $\langle U\rangle_{\chi}$ (dashed curves) as a function of cubic interactions $\bT$ or $\bB$ for the three cases distinguished by color.
  Data corresponds to the cases with $N=7$ in \figref{fig:binodal_spinodal_diff_fixed_interaction_random_phi_sigma1}  averaged over $\chi\in[-30, 30]$.
  }
  \label{fig:withonlyiij_mean_over_chi_N7} 
\end{figure}

To elucidate how cubic interactions affect phases in liquids with more components, we next vary the component count $N$.
If we again used identical interactions between components, we would often find states with more than $N$ phases, seemingly violating Gibb's phase rule~\cite{Gibbs1876}.
This exceptional result is a consequence of the symmetry in the interaction matrix leading to degenerate states; 
for example, if there are states that enrich two of the $N-1$ interacting components (e.g., see \figref{fig:pie}f), there are ${N-1 \choose 2} = \frac12 (N-1)(N-2)$ choices of the component pair, all leading to states of equal energy, which can coexist.
To avoid this problem and study the generic behavior where at most $N$ phases form,
we thus investigate systems with almost equal interaction by setting a finite standard deviation for the interaction matrices ($\sbT=\sbB=\sigma_{\chi}=1$).
\figref{fig:binodal_spinodal_diff_fixed_interaction_random_phi_sigma1} summarizes the results for three different cases, where binary cubic interactions $\bB$ vary without ternary cubic interactions ($\bT=\sbT=0$, left columns), ternary cubic interactions $\bT$ vary without binary cubic interactions ($\bB=\sbB=0$, middle columns), and all cubic interactions exist with identical mean ($\bT=\bB$, $\sbT=\sbB=1$, right columns).

The phase count $M$ shown in the upper row of \figref{fig:binodal_spinodal_diff_fixed_interaction_random_phi_sigma1} indicates that there is a finite region in parameter space where the homogeneous state is stable and phase separation is impossible for all choices of $\bar{\phi}_i$ and $N$ in all three scenarios.
This region generally shrinks for larger $N$, indicating that adding components promotes phase separation.
However, the region stays finite even in the limit of $N\rightarrow\infty$, so that the homogeneous state is always stable for weakly interacting systems; see red dashed line in \figref{fig:binodal_spinodal_diff_fixed_interaction_random_phi_sigma1}(a)(iii, v, vii) and Appendix.
We also observe that the maximal phase count increases with $N$, in accordance with Gibb's phase rule.
Generally, more phases ($M\approx N$) are expected when binary interactions are sufficiently repulsive ($\chi>0$), whereas two phases form for sufficiently attractive binary interactions.
While this general trend is expected~\cite{zwicker2022evolved}, the cubic interactions modify the picture substantially.

The most significant difference between the three scenarios shows in the dependence of the phase count~$M$ on the strength of the cubic interaction; see \figref{fig:binodal_spinodal_diff_fixed_interaction_random_phi_sigma1}(a).
Consistent with the first result section, we find that attractive binary cubic interactions ($\bB<0$) increase $M$, particularly for large $N$ (left columns).
In contrast, ternary cubic interactions tend to increase $M$ when they are repulsive ($\bT>0$, middle columns).
Interestingly, the joint scenario where both cubic interactions are present seems to be dominated by the ternary interactions since the data shown in the right columns of \figref{fig:binodal_spinodal_diff_fixed_interaction_random_phi_sigma1}(a) resembles the data in the middle columns.
However, this conclusion does not seem justified when considering the region without phase separation (enclosed by black lines), which is largest in the joint scenario.
Taken together, this suggests that ternary cubic interactions dominate the phase behavior over binary cubic interactions.

We next check whether the same conclusions could have been obtained from observing the number~$U$ of unstable modes of the homogeneous state.
The density plots shown in \figref{fig:binodal_spinodal_diff_fixed_interaction_random_phi_sigma1}(b) roughly resemble the diagrams in the upper row and particularly capture the increased phase count for larger quadratic interactions $\chi$.
However, the influence of the cubic interactions is not even captured qualitatively.
For instance, repulsive binary cubic interactions ($\bB>0$, left columns) tend to slightly increase the phase count $M$, whereas the number of unstable modes $U$ decreases.
Similarly, the effect of attractive ternary cubic interactions ($\bT<0$, middle columns) is not captured very well.
These differences are revealed more prominently in the direct comparison shown in \figref{fig:binodal_spinodal_diff_fixed_interaction_random_phi_sigma1}(c).
This plot reveals that $U+1$ almost always underestimates $M$ by a significant fraction and that this discrepancy tends to increase for larger component counts~$N$.
Interestingly, the disagreement seems to be less severe when both ternary interactions are considered (right column).

To map out the average influence of cubic interactions, we next average the measured phase counts~$M$ and number~$U$ of unstable modes over the analyze range of quadratic interaction parameters~$\chi$.
\figref{fig:withonlyiij_mean_over_chi_N7} summarizes that cubic interactions have a strong influence on both $M$ and $U$, and that $U+1$ generally  underestimates $M$ significantly.
This presentation also highlights that repulsive ternary cubic interactions ($\bT>0$) generally promote phase separation, whereas repulsive binary cubic interactions ($\bB>0$)  increase $M$ only weakly.
In contrast, attractive binary cubic interactions ($\bB<0$) increase $M$ more strongly than any other scenario, whereas attractive  ternary cubic interactions ($\bT<0$) even decrease $M$.
Moreover, \figref{fig:withonlyiij_mean_over_chi_N7} shows that the combined scenario, where both types of cubic interactions have equal strength, is dominated by the ternary cubic interactions, presumably because it contributes more terms to the free energy; see \Eqref{eq:effective_chi}.
This is particularly surprising for attractive interactions, where the phase count of combined scenario lies below either of the individual cases, whereas the number of unstable modes lies between these cases.
Taken together, this quantification highlights that cubic interactions have a strong effect, which is effectively opposite for binary and ternary cubic interaction.
However, we have so far only considered cases where the interactions between components were relatively similar (small variance), but realistic interactions might vary widely.

\subsection{Variance of binary cubic interactions raises phase count more than variance of ternary interactions}
\begin{figure*}
  \centering
  \includegraphics[width=1\textwidth]{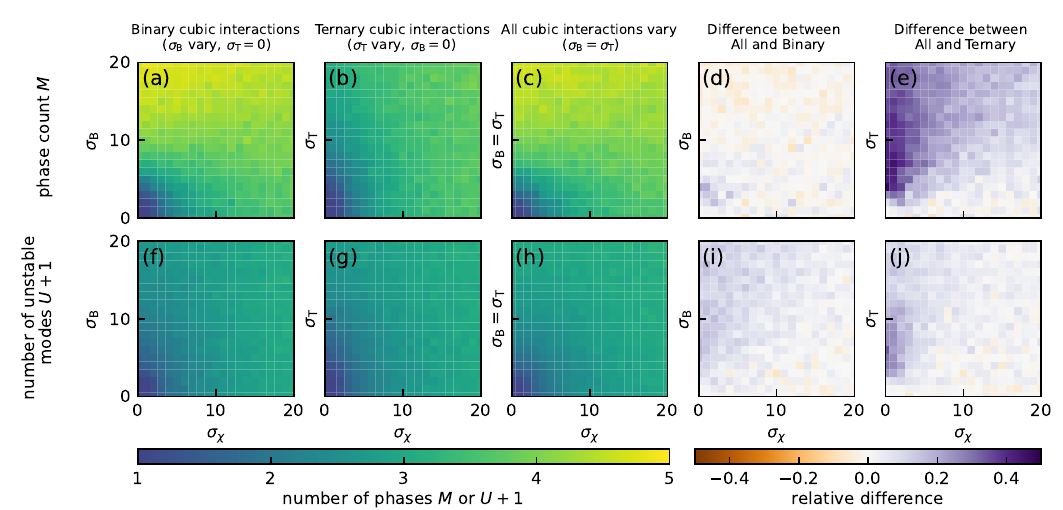}
  \caption{ 
  \textbf{Variance of binary cubic interactions strongly promotes phase count.}
  (a) Mean phase count $M$ as a function of the standard deviations $\sigma_\chi$ and $\sbB$ of binary cubic interactions ($\sbT=0$).
  (b) $M$ as a function of $\sigma_\chi$ and $\sbT$ of ternary cubic interactions ($\sbB=0$).
  (c) $M$ as a function of $\sigma_\chi$ and $\sbB=\sbT$.
  (d) Relative difference~$(\MA-\MB)/\MA$ between $M$ of all cubic interactions, $\MA$, and binary cubic interactions, $\MB$, as a function of $\chi_\sigma$ and $\sbB$.
  (e) $(\MA-\MT)/\MA$ between $\MA$ and the phase count of ternary cubic interactions, $\MT$, as a function of $\chi_\sigma$ and $\sbT$.
  (f)--(j) Number~$U$ of unstable modes for the scenarios corresponding to panels a--e.
  (a)--(j) Additional model parameters are $N=7$ and $\chi=\bB=\bT=0$.
  }
  \label{fig:scaled_sigma} 
\end{figure*}

To study the role of diverse interactions, we next vary the variances $\sigma^2$, $\sbT^2$, and $\sbB^2$ of the random interaction matrices, but we consider vanishing mean interactions ($\chi=\bT=\bB=0$) for simplicity.
\figref{fig:scaled_sigma} shows that the average phase count~$M$ always increases for larger variance~$\sigma_\chi^2$ of the pairwise interactions, consistent with literature~\cite{zwicker2022evolved,Shrinivas2021}.
Similarly, $M$ increases when the variance $\sbB$ of binary cubic interactions, $\sbT$ of ternary cubic interactions, or both are increased.
However, while large values of $\sigma_\chi^2$ lead to roughly $\frac N2$ phases~\cite{Shrinivas2021}, large values of $\sbB^2$, but not $\sbT^2$, can apparently induce many more phases, consistent with the important role of binary cubic interactions revealed in the previous sections.

The combined scenario shown in \figref{fig:scaled_sigma}(c) reveals that the variance of binary cubic interactions dominates the phase count, which is particularly visible in the  direct comparisons shown in panels (d) and (e).
This difference might originate from the strong influence of attractive binary cubic interactions, see \figref{fig:withonlyiij_mean_over_chi_N7}, which will occur frequently in random interactions with vanishing mean.
Interestingly, the difference we observe when analyzing $M$ is not visible if we instead quantify the number~$U$ of unstable modes.
The data shown in panels (f)--(j) again emphasizes that linear stability analysis is not well suited for predicting phase counts.
Taken together, our data suggests that the variance of binary cubic interactions is more crucial than that of ternary interactions, whereas we found the opposite trend when analyzing the mean interactions in the previous section.

\begin{figure*}
  \centering
  \includegraphics[width=1\textwidth]{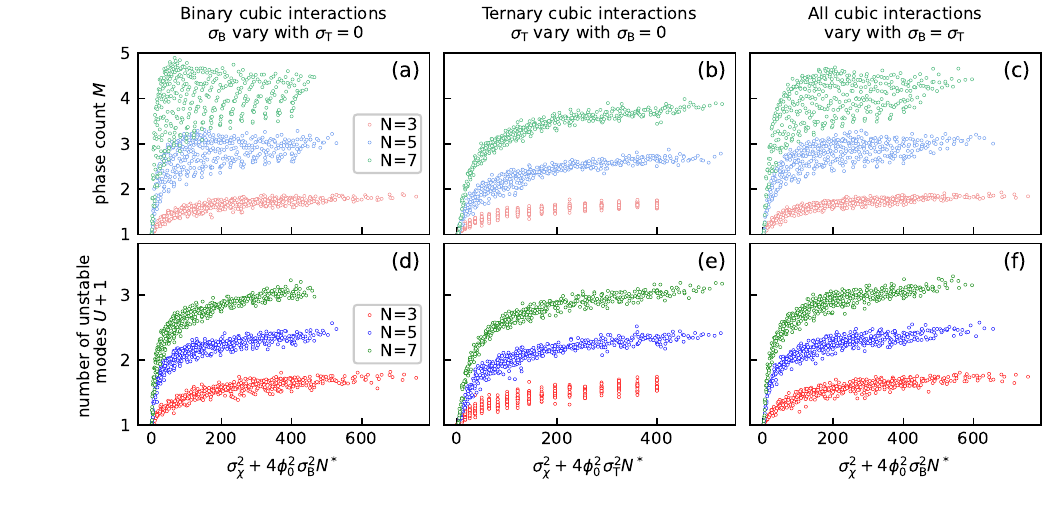}
  \caption{ 
  \textbf{Random cubic interactions can sometimes be reduced to random quadratic interactions.} 
  (a) Mean phase count $M$ as a function of the scaled variance $\sigma^2_\chi+4\phi_0^2\sbB^2N^*$ for various component counts~$N$ ($\sbT=0$, $N^*=2$).
  (b) $M$ as a function of the scaled variance $\sigma^2_\chi+4\phi_0^2\sbT^2N^*$ for various $N$ ($\sbB=0$, $N^*=N-3$).
  (c) $M$ as a function of the scaled variances for $\sbT=\sbB$ ($N^*=N-1$).
  (d)--(f) Number~$U$ of unstable modes as a function of the scaled variances corresponding to the data in panels a--c.
  (a)--(f) Additional model parameters are $\chi=\bB=\bT=0$.
  }
  \label{fig:random_scaling} 
\end{figure*}

The phase count $M$ and the number of unstable modes $U$ shown in \figref{fig:scaled_sigma} suggest a simple dependence on the variances $\sigma_\chi^2$ and $\sbT^2$.
Indeed, \Eqref{eq:effective_chi} indicates that cubic interactions can be absorbed into effective pairwise interactions when analyzing the symmetric homogeneous state where $\bar\phi_i=1/N$; see Appendix.
The pairwise interactions of this reduced system are then a random matrix with zero mean and variance $\sigma^2_\mathrm{eff}=\sigma^2_{\chi}+4\phi_0^2\sbT^2N^*$, where $N^*$ quantifies the number of terms that stem from cubic interaction ($N^*=N-3$ for $\sbB=0$, $N^*=2$ for $\sbT=0$, and $N^* = N-1$ for $\sbB=\sbT$). 
This reduction suggests that systems with the same effective variance $\sigma^2_\mathrm{eff}$ should exhibit similar behavior.
Indeed, Figs.~\ref{fig:random_scaling}(d--f) show that all values $U$ collapse on a line for various $N$ when plotted as a function of $\sigma^2_\mathrm{eff}$.
Such a collapse is also visible for the phase count~$M$ for solely ternary cubic interactions \figref{fig:random_scaling}(b).
In contrast, $M$ does not collapse in systems with binary cubic interactions, and the spread even increase for larger $N$; see panels (a) and (c).
These results are consistent with the previous finding that cubic interactions matter more when binary cubic terms are present.

We also briefly tested whether the dependence on the component count~$N$ can be captured by rescaling $M$ and $U+1$ with $N$.
\figref{fig:random_scaling_rescaled} in the Appendix indicates that the phase count $M$ generally scales with $N$, whereas the number~$U$ of unstable modes does less so.
Since these deviations become significant at large component count ($N>10$), the phase count might also show deviations for larger $N$, which is unfortunately intractable numerically.
Taken together, this analysis demonstrates that random cubic interactions can be described by re-scaled pairwise interactions when only the stability of the homogeneous state is considered.
In contrast, they need to be included explicitly to describe coexisting phases, particularly when binary cubic interactions are present.

We systematically investigated the role of higher-order interactions in multicomponent phase separation. 
Our results demonstrate that cubic interactions influence phase separation significantly by altering the number of phases that coexist and their composition.
By analyzing the coexisting phases in equilibrium, and not just the number of unstable modes of the homogeneous state, we showed that cubic interactions do not merely rescale binary interactions, which was previously suggested for random interactions~\cite{sear2005cytoplasm}.
Moreover, we clearly identified two distinct classes of cubic interactions: 
Cubic interactions between three distinct species tend to have a similar effect to binary interactions, although they can lead to more phases even when the solvent is inert.
In contrast, cubic interactions between only two species, which can be interpreted as composition-dependent binary interactions, have profoundly different influences:
Here, attractive interactions tend to increase the number of phases (whereas increasing the phase count otherwise would typically require repulsion). 
This effect is so strong that it dominates the phase count for random matrices with vanishing mean interactions, which is a case that is often studied.
Taken together, we have shown that cubic interactions have profound effects and that there are different classes, which need to be treated separately.
These conclusions are similar to the effect of higher-order interactions in ecology~\cite{kleinhesselink2022detecting} and will likely translate to quartic interactions and beyond.

Our comparison of the actual phase count in equilibrium to the predictions from a linear stability of the homogeneous state suggests that the latter systematically underestimates the phase count.
The predicted values can deviate by more than $50\%$ and becomes often worse when cubic interactions are significant.
This suggests that linear stability analysis is not very suitable to analyze the phase behavior, particular in complex situations.

Higher-order interactions emerge naturally in complex systems, particularly in biological phase separation.
The interaction parameters, quantified by $b_{ijk}$, can in principle be calculated rigorously from first principles and validated through simulations \cite{bruns1997third, russ2002three, holehouse2023molecular, pappu2023phase}. 
Alternatively, they could be determined by comparison with multi-component phase diagrams, like those presented in \cite{arter2022biomolecular,riback2020composition}. 
Our work focused on random interactions to keep the number of independent parameters manageable, but it is unclear whether random interactions represent realistic systems faithfully since designed interactions provide a more diverse behavior \cite{mao2019phase, chen2023programmable}, and evolutionary optimization can for instance tune the condensate count~\cite{zwicker2022evolved}.
Moreover, allostery could be used to engineer specific cubic interactions~\cite{mccullagh2024allosteric}.
The resulting complex phase diagrams provide many phase boundaries, which could be used to respond to external changes in biological contexts.
In this case, the precise composition of the system will affect what phases form~\cite{thewes2023composition}.
Higher-order interactions expand the range of possibilities of such biological computations, and might thus facilitate the retrieval of target structures in multicomponent liquids~\cite{Teixeira2023}.
These properties reminiscent of information processing might explain why phase separation is ubiquitous in biology.

\section*{acknowledgement}
We thank  Pablo Sartori, Izaak Neri, and Sonali Priyadarshini Nayak for helpful discussions and critical reading of the manuscript. CL thanks Baoshuang Shang and Yuanchao Hu for helpful discussions. We gratefully acknowledge funding from the Max Planck Society and the European Union (ERC, EmulSim, 101044662).

\bibliography{main}

\onecolumngrid
\appendix
  \renewcommand{\thefigure}{S\arabic{figure}}
  \setcounter{figure}{0}

  \section{Theory for multi-component mixtures with higher-order interactions}
  We first write the free energy for the incompressible $N$ component mixtures with internal energies $w_i$ for species $i$, pair interactions $\chi_{ij}$ between species $i$ and $j$, and higher-order interactions. We denote the $h$-order interaction among species $i_1, i_2, \hdots, i_h$ as $B^{(h)}_{i_1,\hdots,i_h}$, and include interactions up to order $H$. The free energy density reads
  \begin{multline}
      f\left[\{\phi_i\}_{i=1,\hdots,N-1}\right] =\frac{k_\mathrm{B}T}{\nu}\Bigg[
       \sum_{i=1}^{N-1} w_i\phi_i
      +\sum_{i=1}^{N-1}\sum_{j=1}^{N-1}\frac12\chi_{ij} \phi_i\phi_j
      +\sum_{h=3}^{H}\frac{1}{h}\sum_{i_1=1}^{N-1}\hdots\sum_{i_h=1}^{N-1}B^{(h)}_{i_1,\hdots,i_h}\phi_{i_1}\hdots\phi_{i_h}
      \\
      +\sum_{i=1}^{N-1}\phi_i\ln\phi_i+\phi_s\ln\phi_s \Bigg],
  \end{multline}
  where $\phi_s=1-\sum_{i=1}^{N-1}\phi_i$ is the volume fraction of the solvent and the last two terms in the square bracket capture translational entropy. If we use $H=3$ and neglect the internal energies that do not change the equilibrium states, this gives Eq.~(1) in the main text. 
  
  The exchange chemical potentials are
  \begin{eqnarray}
    \label{eq:chemical_potential}
      \mu_i =\frac{\nu}{k_BT}\frac{\delta f}{\delta\phi_i}=
      w_i
      +\sum_{j=1}^{N-1}\chi_{ij} \phi_j
      +\sum_{h=3}^{H}\sum_{i_2=1}^{N-1}\hdots\sum_{i_h=1}^{N-1}B^{(h)}_{i,i_2\hdots,i_h}\phi_{i_2}\hdots\phi_{i_h}
  +\ln \phi_i - \ln \phi_s
  \;,
  \end{eqnarray}
  and the osmotic pressure is
  \begin{eqnarray}
    \label{eq:pressure}
    P=\frac{\nu}{k_BT}\left(\sum_{i}^{N-1}\phi_i\mu_i - f\right)=\sum_{i=1}^{N-1}\sum_{j=1}^{N-1}\frac12\chi_{ij} \phi_i\phi_j+ \sum_{h=3}^{H}\frac{h-1}{h}\sum_{i_1}\hdots\sum_{i_h}B^{(h)}_{i_1,\hdots,i_h}\phi_{i_1}\hdots\phi_{i_h}
    -\ln \phi_s
    .
  \end{eqnarray}
  With these expressions, we can in principle predict the equilibrium state for given mean volume fractions $\bar{\phi}_i$ by solving Eq.~(5) in the main text.
  The Hessian matrix~$H_{ij}$ can also be calculated,
  \begin{eqnarray}
    \label{eq:Hessian}
      H_{ij} =\frac{\delta \mu_i}{\delta \phi_j}=
      \chi_{ij}
  +\sum_{h=3}^{H}(h-1)\sum_{i_3}\hdots\sum_{i_h}B^{(h)}_{i,j,i_3,\hdots,i_h}\phi_{i_3}\hdots\phi_{i_h}+\frac{1}{\phi_i}\delta_{ij}+\frac{1}{\phi_s}
  .
  \end{eqnarray}
  In the following, we focus on the system with pair and cubic interactions $b_{ijk}=B^{(3)}_{i,j,k}$. Note we set $\chi_{ij}=\chi_{ji}$ and $b_{ijk}=b_{ikj}=b_{jik}=b_{jki}=b_{kij}=b_{kji}$, as well as $\chi_{ii}=0$ and $b_{iii}=0$, as explained in the main text.
  
  \section{Extra results for identical interactions}
  In this section we expose some results for identical and symmetric interactions, where all off-diagonal terms of pair interactions are $\chi$, all ternary cubic interactions are $\bT$, and all binary cubic interactions are $\bB$. More specifically,
      \begin{align}
        \chi_{ij}=
        \begin{cases}
          \chi\; , & \text{if $i\neq j$} \\
          0\; , & \text{if $i= j$} \\
          \end{cases}
          &&
        b_{ijk}=
      \begin{cases}
      0\; , & \text{if $i=j=k$} \\
      \bT\; , & \text{if $i\neq j$, $j\neq k$, $i\neq k$} \\
      \bB\;, & \text{otherwise}
      \end{cases}
      \; .
      \end{align}
  The free energy density after neglecting the internal energies can then be written as
  \begin{equation}
      f\left[\{\phi_i\}_{i=1,\hdots,N-1}\right] =\frac{k_\mathrm{B}T}{\nu}\Bigg[
      u\left[\{\phi_i\}_{i=1,\hdots,N-1}\right]
      +\sum_{i=1}^{N-1}\phi_i\ln\phi_i+\phi_s\ln\phi_s \Bigg],
  \end{equation}
  where the enthalpic term reads
  \begin{equation}
    \label{eq:energy}
    u\left[\{\phi_i\}_{i=1,\hdots,N-1}\right] =
   \sum_{i<j} \chi \phi_i\phi_j+2\sum_{i<j<k}^{N-1}\bT\phi_i\phi_j\phi_k+\bB\sum_{i<j}\left( \phi_i^2\phi_j+\phi_i\phi_j^2\right).
  \end{equation}
  Absorbing the binary cubic terms to the pair interactions, we have  
  \begin{equation}
    u\left[\{\phi_i\}_{i=1,\hdots,N-1}\right] =
    \sum_{i<j} \chiB_{ij} \phi_i\phi_j+2\sum_{i<j<k}^{N-1}\bT\phi_i\phi_j\phi_k,
   \end{equation}
   where we define a new effective pair interaction for all binary interactions
  \begin{eqnarray}
    \chiB_{ij}=\chi+\bB(\phi_i+\phi_j).
  \end{eqnarray}
  For the special case $N=3$, we further obtain that $\chiB_{12}=\chi+\bB(1-\phiS)$, which is the same as the $\chiEff_{12}$ we have used in the first results section in the main text.
  On the other hand, \Eqref{eq:energy} can be rewritten as 
  \begin{eqnarray}
  &u\left[\{\phi_i\}_{i=1,\hdots,N-1}\right]
    &=\sum_{i<j} \chi \phi_i\phi_j+2\sum_{i<j<k}\bT\phi_i\phi_j\phi_k+\bB\sum_{i<j}\left( \phi_i\left(1-\sum_{l\neq i,l\neq j}\phi_l-\phi_j-\phi_s\right)\phi_j+\phi_i\phi_j^2\right)
    \nonumber\\
    &&=\sum_{i<j} (\chi+\bB) \phi_i\phi_j+2\sum_{i<j<k}\bT\phi_i\phi_j\phi_k-\bB\sum_{i<j} \phi_i\sum_{l\neq i,l\neq j}\phi_l\phi_j-\bB\sum_{i<j} \phi_i\phi_j\phi_s
    \nonumber\\
    &&=\sum_{i<j} (\chi+\bB) \phi_i\phi_j+2\sum_{i<j<k}\bT\phi_i\phi_j\phi_k-3\bB\sum_{i<j<l} \phi_i\phi_l\phi_j-\bB\sum_{i<j} \phi_i\phi_j\phi_s
    \nonumber\\
    &&=\sum_{i<j} (\chi+\bB) \phi_i\phi_j-(3\bB-2\bT)\sum_{i<j<k}\phi_i\phi_j\phi_k-\bB\sum_{i<j} \phi_i\phi_j\phi_s.
  \end{eqnarray}
  If $\chi=-\bB$ and $\bB=\bT$, the equation above can be further simplified to 
  \begin{eqnarray}
  u\left[\{\phi_i\}_{i=1,\hdots,N-1}\right] &=
  -\bB\sum_{i<j<k}\phi_i\phi_j\phi_k-\bB\sum_{i<j} \phi_i\phi_j\phi_s,
  \end{eqnarray}
  which means all components including solvent are equivalent. This can cause the degeneracy and hence more phases as we mentioned in the first result section in the main text.
  
  \subsection{Identical mean volume fractions}
\begin{figure*}
    \centering
    \includegraphics[width=1.0\textwidth]{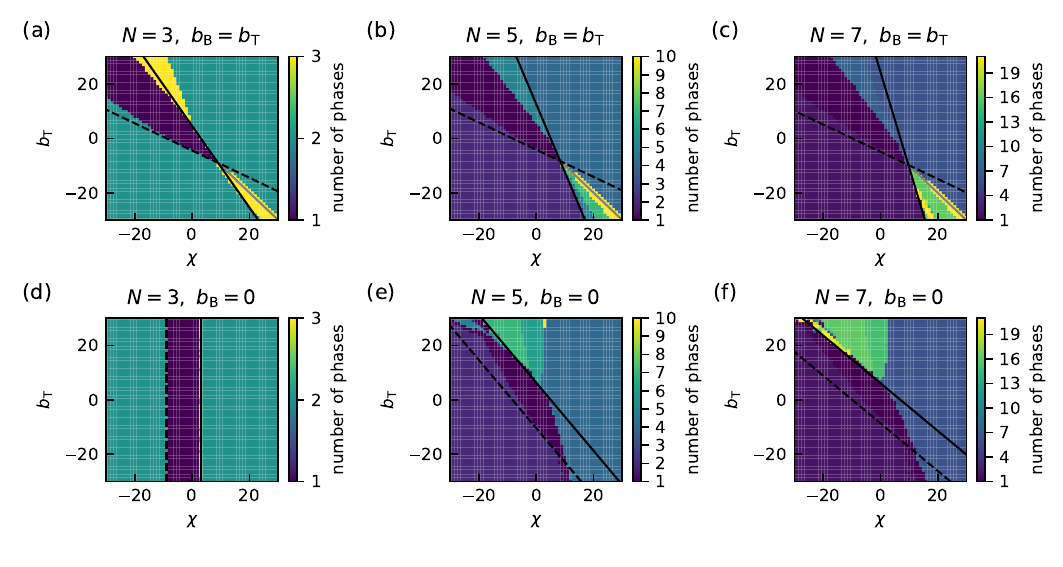}
    \caption{ 
    \textbf{Number of phases for identical interactions with $\bar{\phi}_i=1/N$. }The solid and dashed curves represent $\chi_+$ (\Eqref{eq:chiplus}) and $\chi_-$ (\Eqref{eq:chiminus}), respectively. The grey curve represents $\chi=-\bT$.  }
    \label{fig:binodal_fix} 
  \end{figure*}

\begin{figure*}
    \centering
    \includegraphics[width=1.0\textwidth]{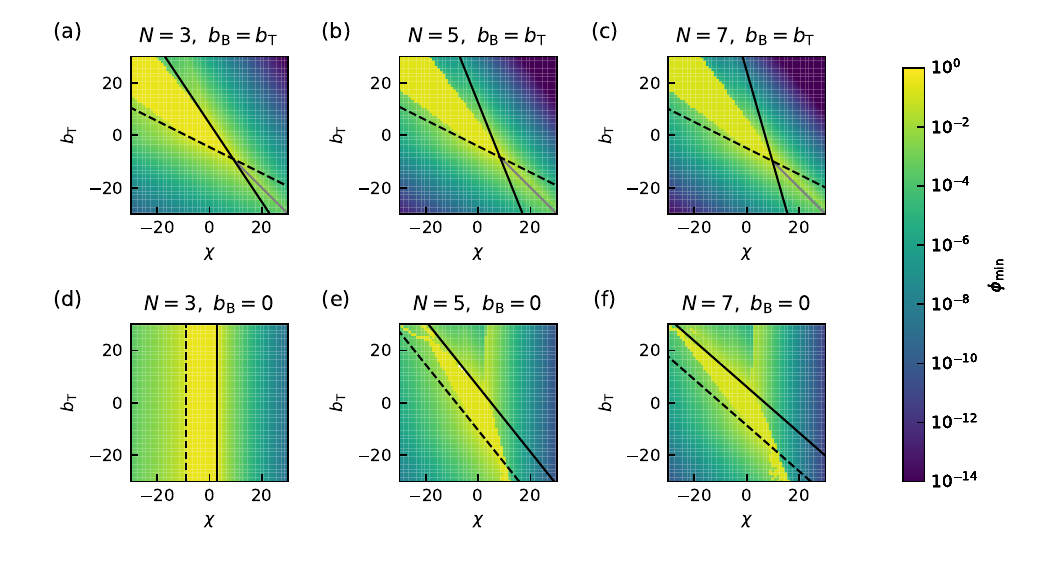}
    \caption{ 
    \textbf{Minimum volume fraction for identical interactions with $\bar{\phi}_i=1/N$.} (a-c) $\phi_{min}=\phi_{l}$ is the lowest volume fraction for solute. (d-f) $\phi_{min}=\phi_{sl}$ is the lowest volume fraction for solvent.}
    \label{fig:phimin} 
  \end{figure*}

   We first consider the special case where all species have the same volume fraction $\bar{\phi}_i=1/N\equiv{\phi}_0$. We numerically solve Eqs.(4) and (5) in the main text and plot the number of species at different $\chi$ and $\bT$ in \figref{fig:binodal_fix}. The minimum volume fraction for all phases and species is plotted in \figref{fig:phimin}.

  When all volume fractions are identical we can analytically solve the eigenvalues of the Hessian matrix to find the instability criteria.
  For $i=j$, 
  \begin{align}
    H_{ii}=2(N-2)\bB\phi_0 +\frac{1}{\phi_0}+\frac{1}{\phi_s}\equiv h_d.
  \end{align}
  For $i\neq j$, 
  \begin{align}
    H_{ij}=\chi + 2\left(2\bB+(N-3)\bT\right)\phi_0 +\frac{1}{\phi_s}\equiv h_o.
  \end{align}
  The eigenvalues are 
  \begin{subequations}
  \begin{align}
    \lambda_+&=h_d-h_o = 2\left((N-4)\bB-(N-3)\bT\right)\phi_0 +\frac{1}{\phi_0}-\chi\; ,
    \\
    \lambda_-&=h_d+(N-2)h_o=2\left((3N-6)\bB+
    (N-2)(N-3)\bT
    \right)\phi_0 +\frac{1}{\phi_0}+(N-1)\frac{1}{\phi_s}+(N-2)\chi
    \; .
  \end{align}
  \end{subequations}
  Solving equations $\lambda_\pm(\chi_\pm)=0$ we obtain 
  \begin{subequations}
  \begin{align}
    \chi_+&=2\left((N-4)\bB-(N-3)\bT\right)\phi_0 +\frac{1}{\phi_0}\; ,
    \\
    \chi_-&=-\frac{2\left((3N-6)\bB+
    (N-2)(N-3)\bT
    \right)\phi_0 +\frac{1}{\phi_0}+(N-1)\frac{1}{\phi_s}}{N-2}
    \; .
  \end{align}
  \end{subequations}
  Using $\phi_0=1/N$ we have 
  \begin{subequations}
  \begin{align}
    \label{eq:chiplus}
    \chi_+&=\frac{-2 (N-3) \bT+2 (N-4) \bB+N^2}{N}
    \\
    \label{eq:chiminus}
    \chi_-&=-\frac{2 (N-3) (N-2) \bT+6 (N-2) \bB+N^3}{(N-2) N}.
  \end{align}
  \end{subequations}
  In \figref{fig:binodal_fix}, \figref{fig:phimin}, and Fig. 1(b) and Fig. 3(b) in the main text, the solid (dashed) black curve represents $\chi_+$ ($\chi_-$).
  When $N\rightarrow\infty$ they become
  \begin{subequations}
  \begin{align}
    \chi_+&=-2 \bT+2 \bB+N
    \\
    \chi_-&=-2 \bT-2-N \;.
  \end{align}
  \end{subequations}
  It is clear that $\bB$ can affect $\chi_+$ but does not appear in $\chi_-$.  From \figref{fig:binodal_fix} we can see that the critical curves (black curves) predicted from linear stability analysis cannot capture the real transition when the three-body interactions are considered.
  
\begin{figure*}
    \centering
    \includegraphics[width=1.0\textwidth]{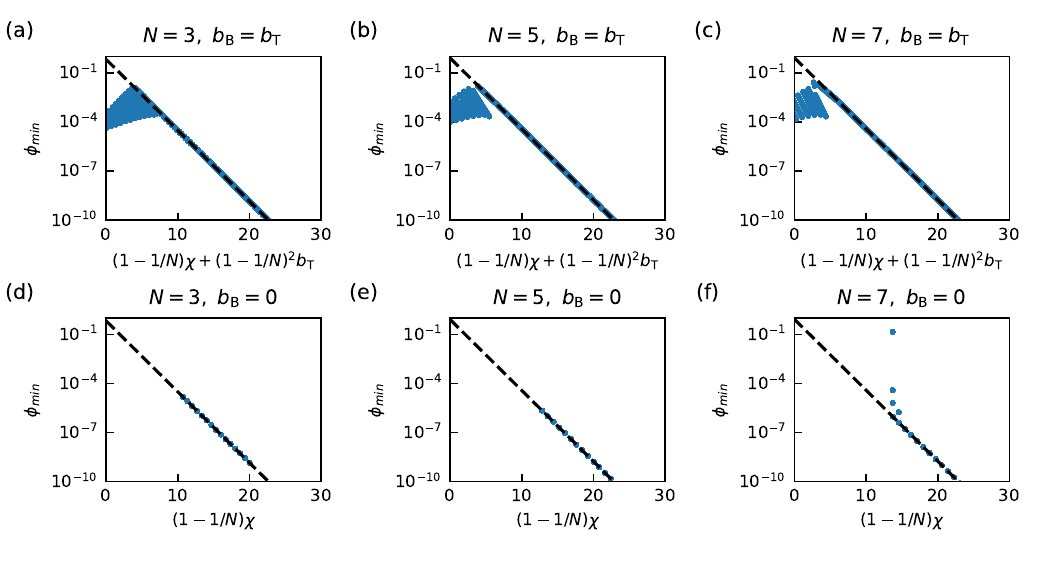}
    \caption{ 
    \textbf{Minimum volume fraction for identical interactions with $\bar{\phi}_i=1/N$ as a function of the scaled interaction for strong repulsion.} Test \Eqref{eq:phi_l}. The dashed curve represents $y=(1-1/N)\exp(-x)$.}
    \label{fig:xplusb_positive} 
  \end{figure*}

  In this special case that all species have the same volume fraction $\bar{\phi}_i=1/N$, Eqs.(4) and (5) in the main text can be solved analytically for large interactions ($|\chi|\gg 1$, $|\bT|\gg 1$, and $|\bB|\gg 1$). 
  % We consider symmetric $\chi$ matrix and $b$ tensor and at conditions $\bar{\phi}_i=1/N=\bar{\phi}_s$. 
  For large positive interactions, we know there are $N-1$ phases. The volume fractions in one phase can be written as $(\phi_h,\phi_l, \hdots, \phi_l, \phi_s)$ where the high fraction $\phi_h\sim 1$ and $N-2$ low fractions $\phi_l\ll 1$. In the other $N-2$ phases the volume fractions can be obtained by exchanging the $i$-th $\phi_l$ with $\phi_h$. 
  The chemical potentials
  $\mu_1 |_{\phi_1=\phi_h}$ and  $\mu_2 |_{\phi_2=\phi_l}$  are
  % \begin{eqnarray}
  %   \partial G/\partial \phi_h = \partial F/\partial \phi_1 |_{\phi_1=\phi_h}
  % \end{eqnarray}
  % and
  % \begin{eqnarray}
  %   \partial G/\partial \phi_l = \sum_{i=2}^{N-1}\partial F/\partial \phi_i |_ {\phi_i=\phi_l} =(N-2)\partial F/\partial \phi_2 |_ {\phi_2=\phi_l}
  % \end{eqnarray}
  % Therefore
  % \begin{eqnarray}
  %   (N-2)\partial G/\partial \phi_h=\partial G/\partial \phi_l
  % \end{eqnarray}
  \begin{subequations}
  \begin{align}
    \mu_1 &=
    \chi \phi_l(N-2)
    +2\bB (N-2) \phi_h\phi_l+\bT(N-2)(N-3)\phi_l^2+\bB(N-2)\phi_l\phi_l
    +\ln \phi_h - \ln \phi_s
  \;,
  \intertext{and}
    \mu_2 &=\chi(\phi_h+(N-3)\phi_l)
    +\bB\phi_h\phi_h+2\bB\phi_h\phi_l+2(N-3)\bT\phi_h\phi_l
    \notag\\
    &\quad
    +2\bB(N-3)\phi_l\phi_l
  +\bT(N-3)(N-4)\phi_l\phi_l+\bB(N-3)\phi_l\phi_l
  +\ln \phi_l - \ln \phi_s
  \;,
  \end{align}
  \end{subequations}
  % \begin{eqnarray}
  %     \mu_1 =
  %     \chi \phi_l(N-2)
  %     +b\left(2(N-2)\phi_h\phi_l+(N-2)(N-2)\phi_l^2\right)
  %     +\ln \phi_h - \ln \phi_s
  % \;,
  % \end{eqnarray}
  % and 
  % \begin{eqnarray}
  %     \mu_2 =\chi(\phi_h+(N-3)\phi_l)
  %     +b\left(\phi_h^2+2(N-2)\phi_h\phi_l+ ((N-2)^2-1)\phi_l^2 \right)
  % +\ln \phi_l - \ln \phi_s
  % \;,
  % \end{eqnarray}
  should satisfy $\mu_1=\mu_2$ and thus we have 
  \begin{eqnarray}
    &&\chi (\phi_h-\phi_l)+\bB\phi_h^2+2(N-3)(\bT-\bB )\phi_h\phi_l
   % \nonumber\\
   % &&
    +((2N-7)\bB-2(N-3)\bT)\phi_l^2+\ln (\phi_l/\phi_h)=0.
  \end{eqnarray}
  Using $\phiS=1/N$, $\phi_l\approx 0$, and $\phi_h=1-\phiS-\phi_l(N-2)\approx 1-\phiS$, we obtain  
  \begin{eqnarray}
    \chi(1-\frac1N)+\bB(1-\frac1N)^2+\ln(\frac{\phi_l}{1-\frac1N})=0.
  \end{eqnarray}
  Hence, %Therefore, we obtain the expression for $\phi_l$, which reads
  % % \begin{eqnarray}
  % %   &&\chi (\phi_h-\phi_l)+\bB\phi_h^2+(2\bB+2(N-3)\bT-2\bB (N-2))\phi_h\phi_l
  % %   \nonumber\\
  % %   &&
  % %   +(2\bB(N-3)+\bT(N-3)(N-4)+\bB(N-3)
  % % - \bT(N-2)(N-3)\phi_l^2-\bB(N-2))\phi_l\phi_l+\ln (\phi_l/\phi_h)=0.
  % % \end{eqnarray}
  % % \begin{eqnarray}
  % %   \chi (\phi_h-\phi_l)+b(\phi_h^2-\phi_l^2)+\ln (\phi_l/\phi_h)=0.
  % % \end{eqnarray}
  % Using $\phi_s=1/N$ and $\phi_h=1-\phi_s-\phi_l(N-2)$ we obtain that
  % \begin{eqnarray}
  %   \chi\left(1-\frac{1}{N}-\phi_l(N-1)\right)+b\left(1-\frac{1}{N}-\phi_l(N-1)\right)\left(1-\frac{1}{N}-\phi_l(N-3)\right)+\ln \left(\frac{\phi_l}{\left(1-\frac{1}{N}-\phi_l(N-2)\right)}\right)=0.
  % \end{eqnarray}
  % Assuming $\phi_l\ll1$, we have 
  % \begin{eqnarray}
  %   \chi\left(1-\frac{1}{N}\right)+b\left(1-\frac{1}{N}\right)\left(1-\frac{1}{N}\right)+\ln \left(\frac{\phi_l}{\left(1-\frac{1}{N}\right)}\right)=0,
  % \end{eqnarray}
  % which gives us the expression
  \begin{eqnarray}
    \label{eq:phi_l} 
    \phi_l=(1-\frac{1}{N})\exp\left(-\chi\left(1-\frac{1}{N}\right)-\bB\left(1-\frac{1}{N}\right)\left(1-\frac{1}{N}\right)\right).
  \end{eqnarray}
  This agrees with our numerical results shown in \figref{fig:xplusb_positive}.
  Note that $\bT$ disappears in the expression of $\phi_l$. If $\bB=0$, $\phi_l$ is determined only by the pair interaction $\chi$ and the number of component $N$. 
  For $N\rightarrow\infty$, we find
  \begin{eqnarray}
    \phi_l=\exp \left(-\chi-\bB\right).
  \end{eqnarray}

  \begin{figure*}
    \centering
    \includegraphics[width=1.0\textwidth]{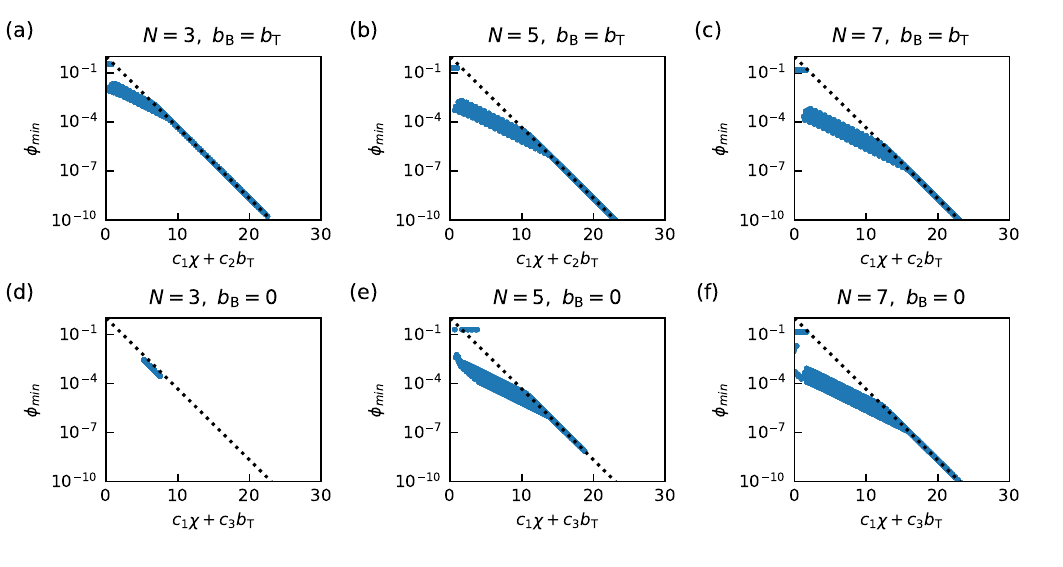}
    \caption{ 
    \textbf{Minimum volume fraction for identical interactions with $\bar{\phi}_i=1/N$ as a function of the scaled interaction for strong attraction.} Test \Eqref{eq:phi_sl}. We use $c_1=(1-1/(N-1))/2$, $c_2=2b(1-1/(N-1)^2)/3$ and $c_3=2b(N-2)(N-3)/(N-1)^2/3$. The dotted curve represents $y=\exp(-x)$.  }
    \label{fig:xplusb_negative} 
  \end{figure*}
  
  % If $b_s=0$, we obtain that
  % \begin{eqnarray}
  %   \mu_1 =
  %   \chi \phi_l(N-2)
  %   +b((N-2)(N-2)-(N-2))\phi_l^2
  %   +\ln \phi_h - \ln \phi_s
  % \;,
  % \end{eqnarray}
  % \begin{eqnarray}
  %   \mu_2 =\chi(\phi_h+(N-3)\phi_l)
  %   +b\left(2(N-3)\phi_h\phi_l+ ((N-3)^2-(N-3))\phi_l^2 \right)
  % +\ln \phi_l - \ln \phi_s
  % \;,
  % \end{eqnarray}
  % which leads to 
  % \begin{eqnarray}
  %  \label{eq:phi_l2} 
  %   \phi_l=(1-\frac{1}{N})\exp\left(-\chi(1-\frac{1}{N})\right)
  % \end{eqnarray}
  %  that is independent of $b$. This expression also agrees with the numerical results, see \figref{fig:xplusb_positive}(d-f).
  
  We next focus on the strong attractive interaction regime. Here we have two phases with fractions $(\phi_l,\hdots,\phi_l,\phi_{sh})$ and $(\phi_h,\hdots,\phi_h,\phi_{sl})$ because of the symmetry of the interactions. Note that $\phi_l(N-1)+\phi_{sh}=1$ and $\phi_h(N-1)+\phi_{sl}=1$.
  We have two equations
  \begin{align}
    \mu_1|_{\phi_1=\phi_l} =\mu_1|_{\phi_1=\phi_h}
    &&
    P|_{\phi_1=\phi_l} =P|_{\phi_1=\phi_h}.
  \end{align}
  Specifically,
  % \begin{eqnarray}
  %   (N-2)\chi\phi_l+((N-1)^2-1)b\phi_l^2+\ln\phi_l-\ln\phi_{sh}=(N-2)\chi\phi_h+((N-1)^2-1)b\phi_h^2+\ln\phi_h-\ln\phi_{sl}
  % \end{eqnarray}
  \begin{multline}
    (N-2)\chi\phi_l+(3(N-2)\bB+(N-2)(N-3)\bT)\phi_l^2+\ln\phi_l-\ln\phi_{sh}
    =\\
    (N-2)\chi\phi_h+(3(N-2)\bB+(N-2)(N-3)\bT)\phi_h^2+\ln\phi_h-\ln\phi_{sl}
  \end{multline}
  and
  % \begin{eqnarray}
  %   \label{eq:pp} 
  %   \frac{1}{2}((N-1)^2-(N-1))\chi\phi_l^2+\frac{2}{3}((N-1)^3-(N-1))b\phi_l^3-\ln\phi_{sh}
  %   \nonumber\\
  %   =\frac{1}{2}((N-1)^2-(N-1))\chi\phi_h^2+\frac{2}{3}((N-1)^3-(N-1))b\phi_h^3-\ln\phi_{sl}
  % \end{eqnarray}
  \begin{multline}
    \label{eq:pp} 
    \frac{1}{2}((N-1)^2-(N-1))\chi\phi_l^2+\frac{2}{3}((N-1)(N-2)(N-3)\bT+3(N-1)(N-2)\bB)\phi_l^3-\ln\phi_{sh}
    =\\
    \frac{1}{2}((N-1)^2-(N-1))\chi\phi_h^2+\frac{2}{3}((N-1)(N-2)(N-3)\bT+3(N-1)(N-2)\bB)\phi_h^3-\ln\phi_{sl}
  \end{multline}
  Sinces the lowest fraction is $\phi_{sl}\ll 1$, we can assume $\phi_h\approx\frac{1}{N-1}$, $\phi_l\approx 0$, and $\phi_{sh}\approx 1$. Hence,
  \begin{subequations}
  \begin{align}
    \label{eq:mul_eq_muh}
   \ln\phi_l &=(N-2)\chi\frac{1}{N-1}+(3(N-2)\bB+(N-2)(N-3)\bT)(\frac{1}{N-1})^2+\ln(\frac{1}{N-1})-\ln\phi_{sl}
  \intertext{and}
   0 &=\frac{1}{2}((N-1)^2-(N-1))\chi(\frac{1}{N-1})^2+\frac{2}{3}((N-1)(N-2)(N-3)\bT+3(N-1)(N-2)\bB)(\frac{1}{N-1})^3-\ln\phi_{sl}.
  \end{align}
  \end{subequations}
  % \begin{eqnarray}
  %   \label{eq:mul_eq_muh}
  %  \ln\phi_l=(N-2)\chi\frac{1}{N-1}+((N-1)^2-1)b(\frac{1}{N-1})^2+\ln(\frac{1}{N-1})-\ln\phi_{sl}
  % \end{eqnarray}
  % and
  % \begin{eqnarray}
  %  0 =\frac{1}{2}((N-1)^2-(N-1))\chi(\frac{1}{N-1})^2+\frac{2}{3}((N-1)^3-(N-1))b(\frac{1}{N-1})^3-\ln\phi_{sl}.
  % \end{eqnarray}
  The latter gives 
  \begin{eqnarray}
    \label{eq:phi_sl}
    \phi_{sl}=\exp \left[\frac{1}{2}\chi (1-\frac{1}{N-1})+\frac{2}{3}\left((1-\frac1{N-1})(1-\frac2{N-1})\bT+3(1-\frac{1}{N-1})\bB\right)\right].
  \end{eqnarray}
  % \begin{eqnarray}
  %   \label{eq:phi_sl}
  %   \phi_{sl}=\exp \left[\frac{1}{2}\chi (1-\frac{1}{N-1})+\frac{2}{3}b(1-\frac{1}{(N-1)^2})\right].
  % \end{eqnarray}
  % Further using \Eqref{eq:mul_eq_muh} we obtain that 
  % \begin{eqnarray}
  %   \phi_l&=&\exp\left((N-2)\chi\frac{1}{N-1}+((N-1)^2-1)b(\frac{1}{N-1})^2+\ln(\frac{1}{N-1})-\ln\left[\frac{1}{2}\chi (1-\frac{1}{N-1})+\frac{2}{3}b(1-\frac{1}{(N-1)^2})\right]\right)
  %   \nonumber\\
  %   &=&\frac{1}{(N-1)\left[\frac{1}{2}\chi (1-\frac{1}{N-1})+\frac{2}{3}b(1-\frac{1}{(N-1)^2})\right]}\exp\left((N-2)\chi\frac{1}{N-1}+((N-1)^2-1)b(\frac{1}{N-1})^2\right).
  % \end{eqnarray}
  % If $b_s=0$, then $\frac{2}{3}((N-1)^3-(N-1))b\phi_i^3$ in \Eqref{eq:pp} is replaced by $\frac{2}{3}(N-1)(N-2)(N-3)b\phi_i^3$, which leads to 
  % \begin{eqnarray}
  %   \label{eq:phi_sl2}
  %   \phi_{sl}=\exp \left[\frac{1}{2}\chi (1-\frac{1}{N-1})+\frac{2}{3}b\frac{(N-2)(N-3)}{{(N-1)^2}}\right].
  % \end{eqnarray}
  % This is independent of $b$ only for $N=3$. 
  % Both \Eqref{eq:phi_sl} and \Eqref{eq:phi_sl2} 
  \Eqref{eq:phi_sl} is consistent with numerical results for sufficient large interactions; see \figref{fig:xplusb_negative}.
  
  \subsection{Ensemble average over different mean volume fractions}
\begin{figure*}
    \centering
    \includegraphics[width=1.0\textwidth]{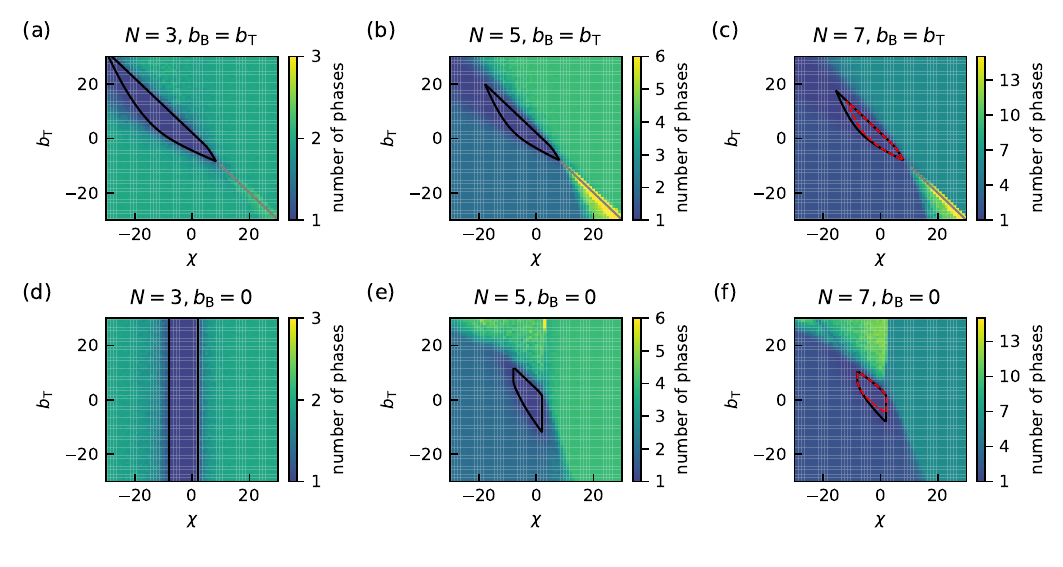}
    \caption{ 
    \textbf{Mean number of phases for different number of components for identical interactions.} The mean number of phases at equilibrium obtained by solving Eq.(4-5) over 100 random mean volume fractions. The black curves are the boundary of the homogeneous stable region that do not separate phase for any fractions. The red dashed curve is the boundary at $N\rightarrow\infty$.  (a)-(c) $\bB=\bT$. (d)-(f) $\bB=0$. The grey curve represents $\chi=-\bT$.  }
    \label{fig:binodal_fixed_interaction_random_phi} 
  \end{figure*}

\begin{figure*}
    \centering
    \includegraphics[width=1.0\textwidth]{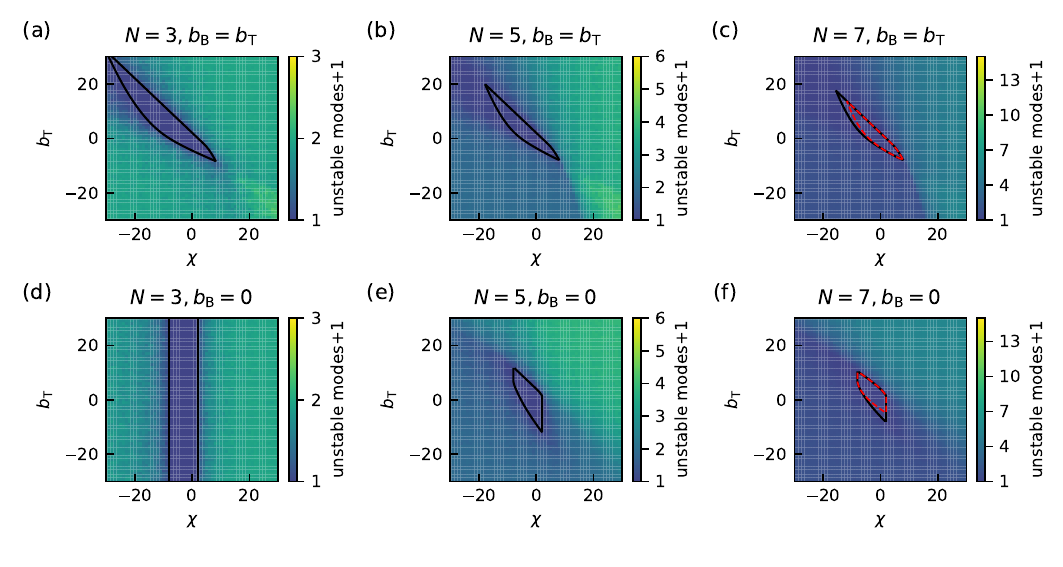}
    \caption{ 
    \textbf{Mean number of unstable modes for identical interactions from linear stability analysis.} For each point $({\chi},{\bT})$, we plot the mean number of unstable modes plus one, i.e., $U+1$, over 100 different mean volume fractions.  (a)-(c) ${\bB}={\bT}$. (d)-(f) ${\bB}=0$. The black and red curves are the same as in \figref{fig:binodal_fixed_interaction_random_phi}.}
    \label{fig:spinodal_fixed_interaction_random_phi} 
  \end{figure*}

\begin{figure*}
    \centering
    \includegraphics[width=1.0\textwidth]{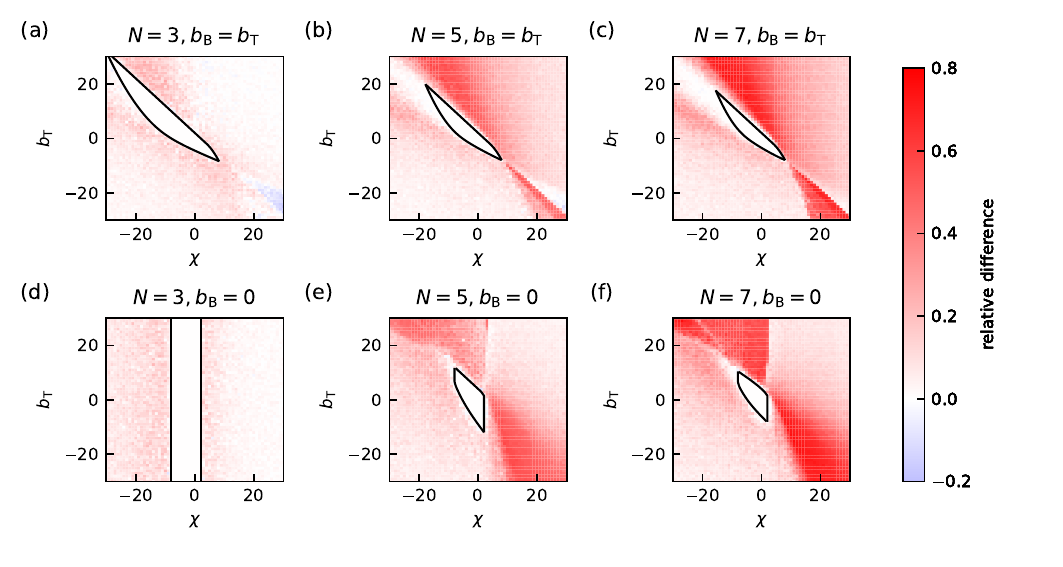}
    \caption{ 
    \textbf{Relative difference of the mean number of phases between spinodal and binodal for identical interactions.} The relative difference $(M-U-1)/M$ where $M$ is the mean number of phases and $U$ is the mean number of unstable modes from linear stability analysis. (a)-(c) $\bB=\bT$. (d)-(f) $\bB=0$. The black curves are the same as in \figref{fig:binodal_fixed_interaction_random_phi}.}
    \label{fig:diff_fixed_interaction_random_phi} 
  \end{figure*}

  In \figref{fig:binodal_fixed_interaction_random_phi} we plot the mean number of unstable modes obtained by sampling 100 different mean volume fractions that follows a uniform distribution~\cite{zwicker2022evolved}.
  In \figref{fig:spinodal_fixed_interaction_random_phi} we plot the mean number of unstable modes. In \figref{fig:diff_fixed_interaction_random_phi} we show the relative different of the mean phase number $M$ and the mean number of unstable modes plus one $U+1$, defined as $(M-U-1)/M$. All are qualitatively the same as the data shown in Fig. 5 in the main text.

  We next explain how we obtain the boundary of the stable region, i.e., the black curves shown in Fig. 2, Fig. 4, and Fig. 5 in the main text. The only assumption is that the phase separation happens first when there are $n$ equivalent species. For a given $N$, the stable region is the intersection of all the stable regions for all possible $n<N$.
  
  \subsubsection{Stable region for $\bB=\bT$}
  For $N=3$ and hence $n=2$, the Hessian is 
  \begin{eqnarray}
    H=\left(
      \begin{array}{cc}
       2 \bB \phi _2+\frac{1}{\phi _s}+\frac{1}{\phi _1} & 2 \bB \left(\phi _1+\phi _2\right)+\frac{1}{\phi _s}+\chi  \\
       2 \bB \left(\phi _1+\phi _2\right)+\frac{1}{\phi _s}+\chi  & 2 \bB \phi _1+\frac{1}{\phi _s}+\frac{1}{\phi _2} \\
      \end{array}
      \right).
  \end{eqnarray}
  All eigenvalues should be larger than $0$ to ensure no phase separation can happen. We assume that $\phi_1=\phi_2$ because of symmetry and use $\phi_1+\phi_2+\phi_s=1$,
  \begin{eqnarray}
    H=\left(
      \begin{array}{cc}
        \bB (1-\phi_s)+\frac{1}{\phi _s}+\frac{2}{1-\phi _s} & 2 \bB \left(1-\phi_s\right)+\frac{1}{\phi _s}+\chi  \\
       2 \bB \left(1-\phi_s\right)+\frac{1}{\phi _s}+\chi    & \bB (1-\phi_s)+\frac{1}{\phi _s}+\frac{2}{1-\phi _s} \\
      \end{array}
      \right).
  \end{eqnarray}
  The eigenvalues are 
  \begin{eqnarray}
    \lambda_\pm=\bB (1-\phi_s)+\frac{1}{\phi _s}+\frac{2}{1-\phi _s} \mp \left(2 \bB \left(1-\phi_s\right)+\frac{1}{\phi _s}+\chi\right),
  \end{eqnarray}
  or more specifically
  \begin{subequations}
  \begin{align}
    \lambda_+&=-\bB(1-\phi_s)+\frac{2}{1-\phi_s}-\chi,
  \\
    \lambda_-&=3\bB(1-\phi_s)+\frac{2}{1-\phi_s}+\chi+\frac{2}{\phi_s}.
  \end{align}
  \end{subequations}
  Simultaneously satisfying $\lambda_\pm>0$ for all possible $\phi_s$ leads to the black curves in Fig.2(a) in the main text. 
  
  Generally, for $n<N$, the two eigenvalues are
  \begin{subequations}
  \begin{align}
    \lambda_+&=\frac{n}{1-\phi_s}-\chi-2\bB\frac{1-\phi_s}{n}
  \\
    \label{eq:lambda_minus}
    \lambda_-&=(n-1)\chi +\frac{n}{\phi_s}+\frac{n}{1-\phi_s}+2\bB(n-1)(n+1)\frac{1-\phi_s}{n}\;.
  \end{align}
  \end{subequations}
  The condition $\lambda_+>0$ leads to 
  \begin{eqnarray}
    \label{eq:chi_less_than}
    \chi<\frac{n}{1-\phi_s}-2\bB\frac{1-\phi_s}{n}\;.
  \end{eqnarray}
   We define $f(\phi_s)=\frac{n}{1-\phi_s}-2\bB\frac{1-\phi_s}{n}$ and need to find the minimum of $f(\phi_s)$. When $\bB>0$, $f$ increases as $\phi_s$ increases, and hence $\min(f)=f(\phi_s=0)$, which is 
  \begin{eqnarray}
    \label{eq:chi_positive_b}
    \chi<n-\frac{2\bB}{n},
  \end{eqnarray}
  which reduces to $\chi<=2-\bB$ since $n=2$.
  When $\bB<0$, however, 
  \begin{eqnarray}
    f'(\phi_s)=\frac{n}{(1-\phi_s)^2}+\frac{2\bB}{n}=0,
  \end{eqnarray}
  which leads to 
  \begin{eqnarray}
    \phi_s^*=1-\sqrt{\frac{n^2}{-2\bB}}.
  \end{eqnarray}
  If $\phi_s^*>0$, i.e., $\bB<-n^2/2$,
  \begin{eqnarray}
    \min(f)=f(\phi_s=\phi_s^*)=\frac{n}{1-(1-\sqrt{\frac{n^2}{-2\bB}})}-2\bB\frac{1-(1-\sqrt{\frac{n^2}{-2\bB}})}{n}=2\sqrt{-2\bB},
  \end{eqnarray}
  and hence 
  \begin{eqnarray}
    \label{eq:chi_negative_b}
    \chi<2\sqrt{-2\bB}.
  \end{eqnarray}
  If $\phi_s^*<0$, then still $\min(f)=f(\phi_s=0)$.
  Taken together, 
  \begin{align}
    \chi < \begin{cases}
    2\sqrt{-2\bB} &  \bB<-n^2/2 \\
    %n-\frac{2\bB}{n}=
    2-\bB & 	\bB>-n^2/2\\
    \end{cases}
  \end{align}
  %\DZ{This what was written earlier:
  %\begin{itemize}
  %  \item if $\bB<-n^2/2$, $\chi<2\sqrt{-2b}$.
  %  \item if $\bB>-n^2/2$, $\chi<n-\frac{2\bB}{n}=2-\bB$.
  %\end{itemize}
  %I don't understand the last inequality. Why is $n-\frac{2\bB}{n}=2-\bB$?
  %}
  Hence, $\lambda_+$ at $n=2$ gives the most strict criteria and hence it is still the boundary even if $n\rightarrow\infty$. For $\lambda_-$, it is more difficult because the minimum depends on both $n$ and $\phi_s$. We know that if $n\rightarrow\infty$ \Eqref{eq:lambda_minus} with $\lambda_->0$ converges to 
  \begin{eqnarray}
    \label{eq:lambda_infty}
    \chi+\frac{1}{\phi_s}+\frac{1}{1-\phi_s}+2\bB(1-\phi_s)>0.
  \end{eqnarray}
  We numerically solve $\lambda_->0$ with \Eqref{eq:lambda_minus} up to $n=1000$ and find it agrees well with \Eqref{eq:lambda_infty}. This large $N$ limit boundary of the stable region is plotted as red dashed line in Fig. 5(vii) in the main text.
  
  \subsubsection{Stable region for $\bB=0$}
  Now we consider $\bB=0$. We obtain the
  two eigenvalues
  \begin{subequations}
  \begin{align}
    \lambda_+&=\frac{n}{1-\phi_s}-\chi-2\bT\frac{n-2}{n}(1-\phi_s),
    \\
    \lambda_-&=\frac{n}{1-\phi_s}+\frac{n}{\phi_s}+(n-1)\chi+2\bT\frac{(n-1)(n-2)}{n}(1-\phi_s).
  \end{align}
  \end{subequations}
  Here, $\lambda_+>0$ leads to 
  \begin{eqnarray}
    \chi<\frac{n}{1-\phi_s}-2\bT\frac{n-2}{n}(1-\phi_s).
  \end{eqnarray}
  This is exactly the same as \Eqref{eq:chi_less_than} if we 
  replace  $\bT(n-2)$ by $\bB$ and hence we obtain
  \begin{align}
    \chi < \begin{cases}
      2\sqrt{-2\bT(n-2)} & \bT(n-2)<-n^2/2 \\
       n-\frac{2\bT(n-2)}{n} & \bT(n-2)>-n^2/2 \\
    \end{cases}
  \end{align}
  %\begin{itemize}
  %  \item if $\bT(n-2)<-n^2/2$, $\chi<2\sqrt{-2\bT(n-2)}$.
  %  \item if $\bT(n-2)>-n^2/2$, $\chi<n-\frac{2\bT(n-2)}{n}$.
  %\end{itemize}
  Similar to the case of $\bB=\bT$, $\lambda_-$ is more complicated but we can numerically obtain the boundary of the stable region, as the black curves shown in Fig. 5(iv--v) in the main text. We also plot the  boundary for $N\rightarrow\infty$ as the red dashed line in Fig. 5(v) in the main text.
  
  \subsubsection{Stable region for $\bT=0$}
  Now we consider $\bT=0$. We obtain the
  two eigenvalues
  \begin{subequations}
  \begin{align}
    \lambda_+&=\frac{n}{1-\phi_s}-\chi+2\bB(1-\phi_s)\frac{n-3}{n},
  \\
    \lambda_-&=\frac{n}{1-\phi_s}+\frac{n}{\phi_s}+(n-1)\chi+6\bB\frac{n-1}{n}(1-\phi_s).
  \end{align}
  \end{subequations}
  Here, $\lambda_+>0$ leads to 
  \begin{eqnarray}
    \chi<\frac{n}{1-\phi_s}+2\bB\frac{n-3}{n}(1-\phi_s).
  \end{eqnarray}
  Similar as before, we obtain the boundary of the stable region; see the black curves shown in Fig. 5(i--iii) in the main text. We also plot the  boundary for $N\rightarrow\infty$ as the red dashed line shown in Fig. 5(iii) in the main text.
  \section{Extra results for random interactions}
  Here we report some results for Gaussian random interactions.
  \subsection{Stability analysis for identical volume fractions}
  We first consider the special case with identical volume fractions $\bar{\phi}_i=\phi_0=1/N$.
  The $(N-1)\times (N-1)$ Hessian matrix in \Eqref{eq:Hessian} becomes
  \begin{align}
    \label{eq:hij_random}
    H_{ij}=\chi_{ij}   
    +\sum_{h=3}^{H}(h-1)\phi_0^{h-2}\sum_{i_3=1}^{N-1}\hdots\sum_{i_h=1}^{N-1}B^{(h)}_{i,j,i_3,\hdots,i_h}+\frac{1}{\phi_0}\delta_{ij}+\frac{1}{\phi_s}.
  \end{align}
  We consider the system only with pair random interactions, i.e., 
  \begin{align}
    \chi_{ij}=
    \begin{cases}
      \mathcal{N}(\chi,\sigma)\; , & \text{if $i\neq j$} \\
      0\; , & \text{if $i= j$} \\
      \end{cases}
      \; ,
  \end{align}
  This is well studied  \cite{Sear2003}. More specifically, there are $N-2$ eigenvalues that follows the semi-circle distribution
  \begin{eqnarray}
    W(x)=\frac{2}{\pi}\sqrt{1-x^2},
  \end{eqnarray}
  if $|x|<1$
  where 
  \begin{eqnarray}
    x=\frac{\lambda+\chi-\frac{1}{\phi_0}}{2\sigma (N-1)^{1/2}},
  \end{eqnarray}
  and $\lambda$ is the eigenvalue. One eigenvalue is asymptotically Gaussian distributed
  \begin{eqnarray}
    P(\lambda)=\frac{1}{\sqrt{2\pi (2\sigma^2)}}\exp\left[-\frac{(\lambda-\frac{1}{\phi_i}+\chi-((N-1)(\chi+\frac{1}{\phi_s})+\frac{\sigma^2}{\chi+\frac{1}{\phi_s}}))^2}{2(2\sigma^2)}\right].
  \end{eqnarray}
  We next add the higher-order interactions and assume that they also follow a normal distribution $B^{(h)}_{i,j,i_3,\hdots,i_h}\sim \mathcal{N}(b_h,\sigma_{h})$. We can construct an effective pair interaction 
  \begin{eqnarray}
    \tilde{\chi}_{ij}^{(H)}=\chi_{ij}  
    +\sum_{h=3}^{H}(h-1)\phi_0^{h-2}\sum_{i_3=1}^{N-1}\hdots\sum_{i_h=1}^{N-1}B^{(h)}_{i,j,i_3,\hdots,i_h},
  \end{eqnarray}
  with the mean value 
  \begin{eqnarray}
    \tilde{\chi}^{(H)}=\chi+\sum_{h=3}^{H}(h-1)\phi_0^{h-2}b_h(N-1)^{h-2},
  \end{eqnarray}
  and the variance
  \begin{eqnarray}
    \label{eq:variance}
    \left(\tilde{\sigma}^{(H)}\right)^2=\sigma^2+\sum_{h=3}^{H}\left((h-1)\phi_0^{h-2}\right)^2\sigma_{h}^2(N-1)^{h-2}
    \;.
  \end{eqnarray}
  Therefore, for identical volume fractions, the higher order interactions are trivial and can be represented by an effective pair interaction. Note that this good feature is based on the assumption that all elements $B^{(h)}_{i,j,i_3,\hdots,i_h}$ follow the same distribution and are independent. In the model we considered in the main text, however, we need to be more careful because not all three-body interaction terms follow the the same normal distribution. For example, when $\sbs=0$, the elements $b_{iij}$ are fixed to $0$. Therefore, we need change the value $N-1$ in \Eqref{eq:variance} by the real number of random elements and we denote it by $N^*$. For $\sbs=0$, $N^*=N-3$; for $\sbs=\sbo$, $N^*\approx N-1$; and for $\sbT=0$, $N^*\approx 2$. Therefore, we rescale the variance in Fig. 8 in the main text, by assuming that this also works for arbitrary mean volume fractions. A more strict way to analytically study this can follow \cite{thewes2023composition}.
  
  \subsection{Numerical results for different volume fractions}
\begin{figure*}
    \centering
    \includegraphics[width=1\textwidth]{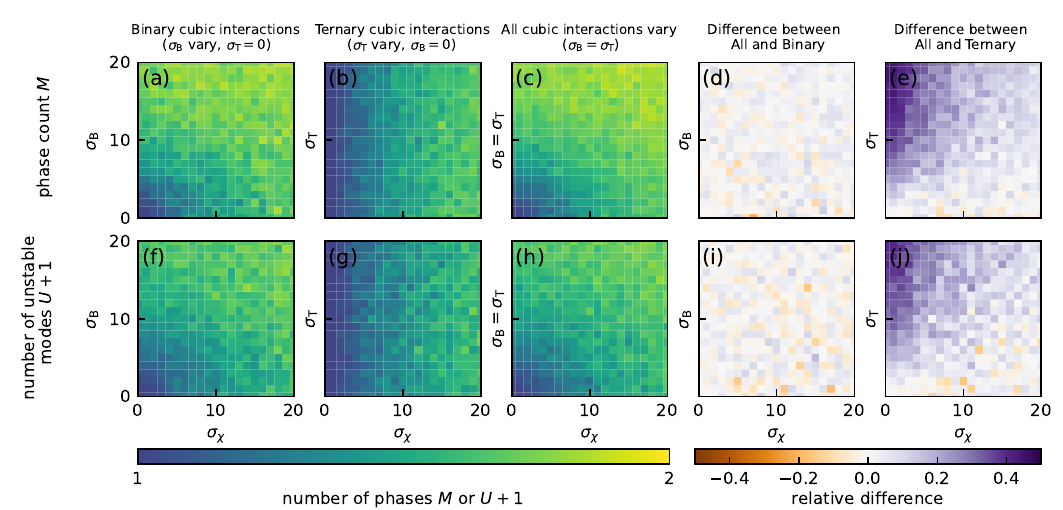}
    \caption{ 
    \textbf{Heterogenous interactions promote many phases for $N=3$.}
    Mean phase count $M$ as a function of the standard deviations $\sigma_\chi$ and (a) $\sbs$ of the pairwise cubic interactions, (b) $\sbo$ ternary cubic interactions, and (c) all cubic interactions with $\sbs=\sbo$, for $N=3$. (d) The relative difference between the phase count of all cubic interactions $\MA$ and binary cubic interactions $\MB$, i.e., $(\MA-\MB)/\MA$. (e) The relative difference between the phase count of all cubic interactions $\MA$ and ternary cubic interactions $\MT$, i.e., $(\MA-\MT)/\MA$.
    (f-h) Number of unstable modes $U+1$ as a function of $\sigma_\chi$ and the corresponding standard deviations. (i-j) Relative difference.
    Model parameters are $\chi=\bs=\bo=0$.
    }
    \label{fig:random_random_binodal_spinodal_diff_N3} 
  \end{figure*}

\begin{figure*}
    \centering
    \includegraphics[width=1\textwidth]{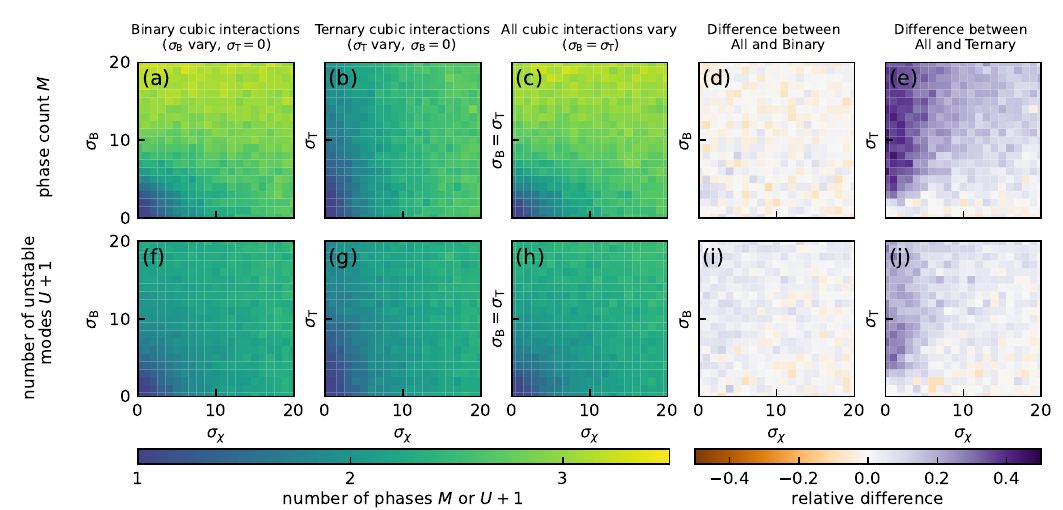}
    \caption{ 
    \textbf{Heterogenous interactions promote many phases for $N=5$.}
    Mean phase count $M$ as a function of the standard deviations $\sigma_\chi$ and (a) $\sbs$ of the pairwise cubic interactions, (b) $\sbo$ ternary cubic interactions, and (c) all cubic interactions with $\sbs=\sbo$, for $N=5$. (d) The relative difference between the phase count of all cubic interactions $\MA$ and binary cubic interactions $\MB$, i.e., $(\MA-\MB)/\MA$. (e) The relative difference between the phase count of all cubic interactions $\MA$ and ternary cubic interactions $\MT$, i.e., $(\MA-\MT)/\MA$.
    (f-h) Number of unstable modes $U+1$ as a function of $\sigma_\chi$ and the corresponding cubic standard deviations. (i-j) The relative difference for the number of unstable modes.
    Additional model parameters are $\chi=\bs=\bo=0$.
    }
    \label{fig:random_random_binodal_spinodal_diff_N5} 
  \end{figure*}

\begin{figure*}
    \centering
    \includegraphics[width=1\textwidth]{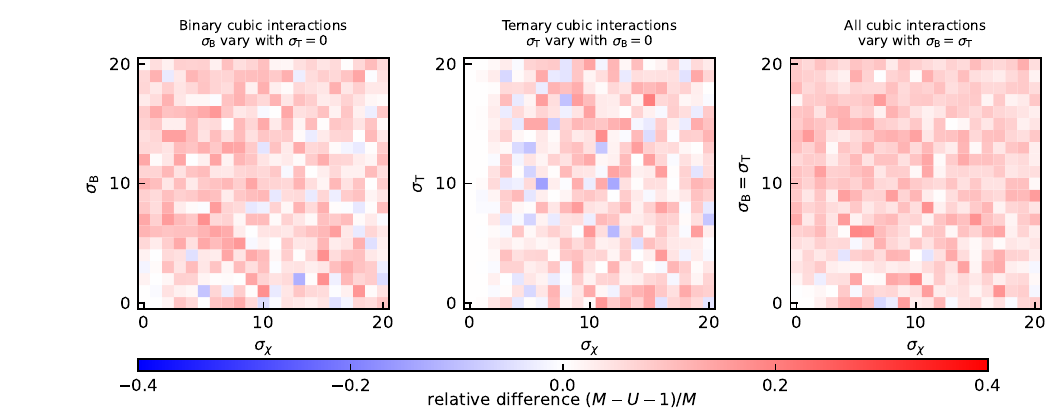}
    \caption{ 
    \textbf{Relative difference between phase count and number of unstable modes for $N=3$ with random interactions and random compositions.}
    Additional model parameters are $\chi=\bs=\bo=0$.
    }
    \label{fig:random_random_diff_N3} 
  \end{figure*}

\begin{figure*}
    \centering
    \includegraphics[width=1\textwidth]{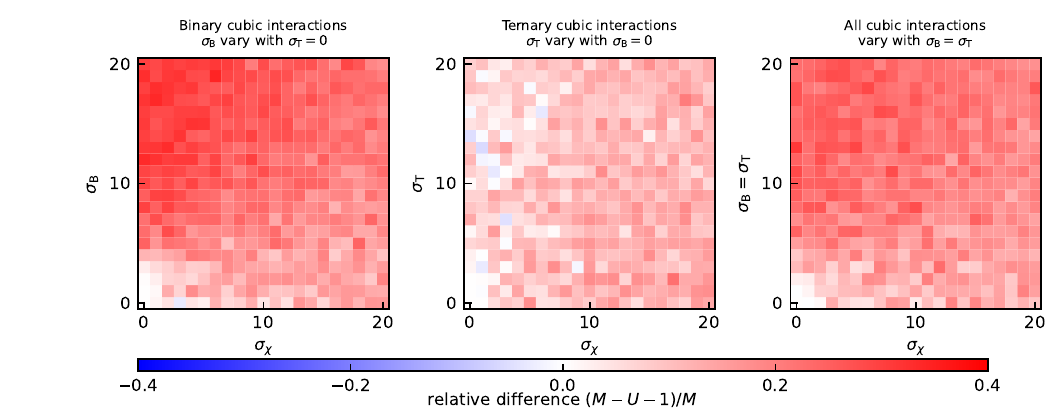}
    \caption{ 
    \textbf{Relative difference between phase count and number of unstable modes for $N=5$ with random interactions and random compositions.}
    Additional model parameters are $\chi=\bs=\bo=0$.
    }
    \label{fig:random_random_diff_N5} 
  \end{figure*}

\begin{figure*}
    \centering
    \includegraphics[width=1\textwidth]{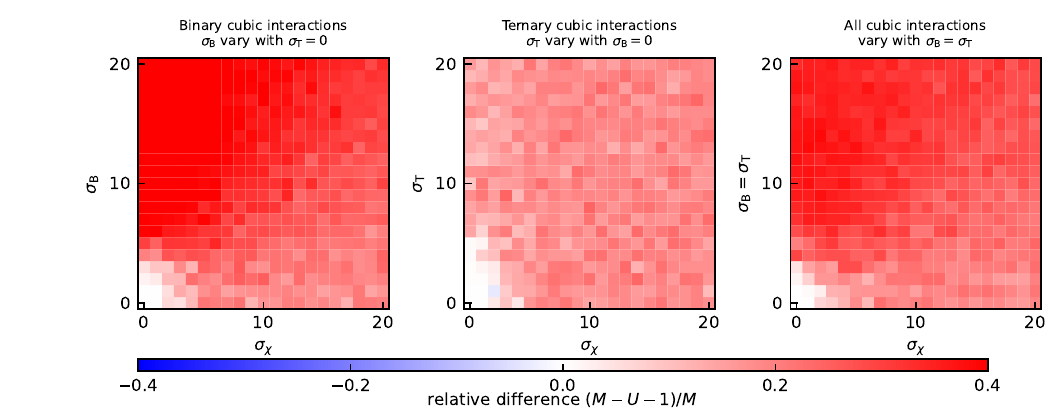}
    \caption{ 
    \textbf{Relative difference between phase count and number of unstable modes for $N=7$ with random interactions and random compositions.}
    Additional model parameters are $\chi=\bs=\bo=0$.
    }
    \label{fig:random_random_diff_N7} 
  \end{figure*}

\begin{figure*}
    \centering
    \includegraphics[width=1\textwidth]{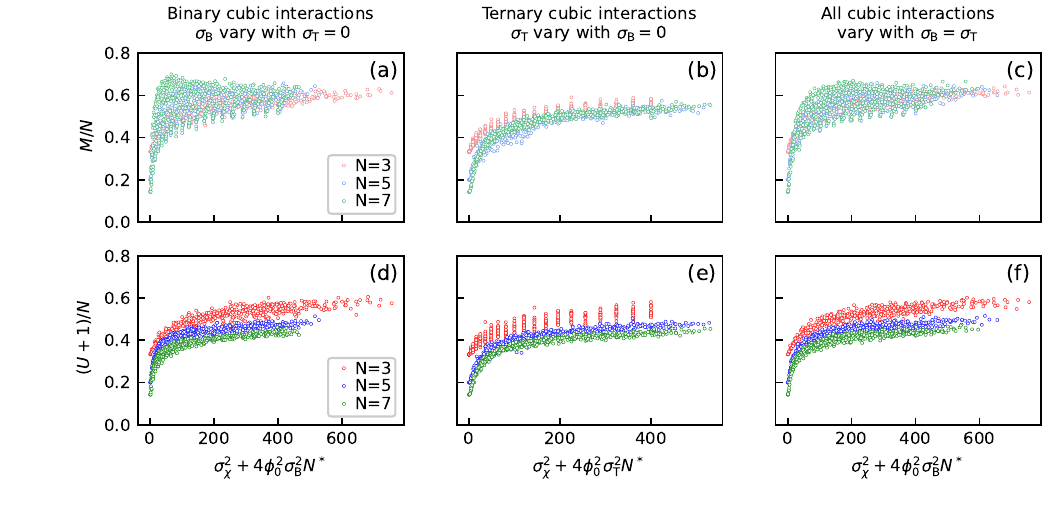}
    \caption{ 
    \textbf{Random cubic interactions can sometimes be reduced to random quadratic interactions.} %{Effective variance can explain the number of unstable modes but not phase count.}
  %  (a--c) Mean phase count $M$ as a function of the scaled variance $\sigma^2_\chi+4\phi_0^2\sbB^2N^*$ or $\sigma^2_\chi+4\phi_0^2\sbT^2N^*$ for various numbers~$N$ of components.
    (a) Scaled mean phase count $M/N$ as a function of the scaled variance $\sigma^2_\chi+4\phi_0^2\sbB^2N^*$ for various component counts~$N$ ($\sbT=0$, $N^*=2$).
    (b) $M/N$ as a function of the scaled variance $\sigma^2_\chi+4\phi_0^2\sbT^2N^*$ for various $N$ ($\sbB=0$, $N^*=N-3$).
    (c) $M/N$ as a function of the scaled variances for $\sbT=\sbB$ ($N^*=N-1$).
    (d)--(f) Scaled number of unstable modes~$(U+1)/N$ as a function of the scaled variances corresponding to the data in panels a--c.
    (a)--(f) Additional model parameters are $\chi=\bB=\bT=0$.
    }
    \label{fig:random_scaling_rescaled} 
  \end{figure*}

   We have sampled $30$ different random interactions and $30$ different mean volume fractions for each point $(\sigma_\chi,\sbT)$. \figref{fig:random_random_binodal_spinodal_diff_N3} and \figref{fig:random_random_binodal_spinodal_diff_N5} are the figure for $N=3$ and $N=5$ respectively, similar to Fig. 7 in the main text.  \figref{fig:random_random_diff_N3}, \figref{fig:random_random_diff_N5} and \figref{fig:random_random_diff_N7} show the differences $(M-U-1)/M$ for $N=3,\ 5,\ 7$, respectively. All results demonstrate that $M>U+1$. 
   \figref{fig:random_scaling_rescaled} shows the phase count scaled by the number of components.

\end{document}